\documentclass{aa}  

\usepackage{natbib}
\usepackage{graphicx}
\usepackage{txfonts}

\usepackage[dvipsnames]{xcolor}
\usepackage{tikz}
\usetikzlibrary{bayesnet}

\usepackage{diagbox}
\usepackage{subcaption}

\newcommand{\de}{\text{d}}
\newcommand{\Msun}{\text{M}_{\odot}}
\newcommand{\kmsec}{\text{km}\,\text{s}^{-1}}

\newcommand{\kpc}{\text{kpc}}
\newcommand{\pc}{\text{pc}}

\newcommand{\Msunppcc}{\Msun\pc^{-3}}

\newcommand{\popp}{\boldsymbol{\Psi}}

\newcommand{\data}{\boldsymbol{d}}

\newcommand{\sigm}{\text{sigm}}

\defcitealias{PaperI}{Paper~I}
\defcitealias{PaperII}{Paper~II}
\defcitealias{PaperIII}{Paper~III}

\usepackage{hyperref}
\begin{document}

   \title{Weighing the Galactic disk using phase-space spirals \\ IV. Tests on a three-dimensional galaxy simulation}
   \titlerunning{Weighing the Galactic disk using phase-space spirals IV}

   \author{A. Widmark
          \inst{1}
          \and
          J. A. S. Hunt
          \inst{2}
          \and
          C. F. P. Laporte
          \inst{3}
          \and
          G. Monari
          \inst{4}
          }

   \institute{Dark Cosmology Centre, Niels Bohr Institute, University of Copenhagen, Jagtvej 128, 2200 Copenhagen N, Denmark\\
   \email{axel.widmark@nbi.ku.dk}
   \and
   Center for Computational Astrophysics, Flatiron Institute, 162 5th Av., New York City, NY 10010, USA
   \and
   Institut de Ci\`encies del Cosmos (ICCUB), Universitat de Barcelona (IEEC-UB), Mart\'i i Franqu\`es 1, 08028 Barcelona, Spain
   \and
   Universit\'e de Strasbourg, CNRS UMR 7550, Observatoire astronomique de Strasbourg, 11 rue de l'Universit\'e, 67000 Strasbourg, France
    }

   \date{Received Month XX, XXXX; accepted Month XX, XXXX}

% \abstract{}{}{}{}{} 
% 5 {} token are mandatory
 
  \abstract{
  In this fourth article on weighing the Galactic disk using the shape of the phase-space spiral, we have tested our method on a billion particle three-dimensional $N$-body simulation, comprised of a Milky Way like host galaxy and a merging dwarf satellite. The main purpose of this work was to test the validity of our model's fundamental assumptions that the spiral inhabits a locally static and vertically separable gravitational potential. These assumptions might be compromised in the complex kinematic system of a disturbed three-dimensional disk galaxy; in fact, the statistical uncertainty and any potential biases related to these assumptions are expected to be amplified for this simulation, which differs from the Milky Way in that it is more strongly perturbed and has a phase-space spiral that inhabits higher vertical energies. We constructed 44 separate data samples from different spatial locations in the simulated host galaxy. Our method produced accurate results for the vertical gravitational potential of these 44 data samples, with an unbiased distribution of errors with a standard deviation of 7~\%. We also tested our method under severe and unknown spatially dependent selection effects, also with robust results; this sets it apart from traditional dynamical mass measurements that are based on the assumption of a steady state and which are highly sensitive to unknown or poorly modelled incompleteness. Hence, we will be able to make localised mass measurements of distant regions in the Milky Way disk, which would otherwise be compromised by complex and poorly understood selection effects.
  }


   \maketitle
%
%________________________________________________________________

\section{Introduction}\label{sec:intro}

An important avenue for learning about the Milky Way is measuring its gravitational potential and matter density using stellar dynamics \citep[e.g.][]{1998MNRAS.294..429D,Widrow:2008yg,2017MNRAS.465...76M,2021MNRAS.506.5721A}. This is especially important for the Galaxy's distribution of dark matter \citep{Read2014,2021RPPh...84j4901D}; for example, the local dark matter density is proportional to the signal strength in direct and some indirect dark matter detection experiments \citep{Jungman:1995df,2015PrPNP..85....1K}. In recent years, the \emph{Gaia} satellite \citep{2018A&A...616A...1G} has revolutionised the research field of Galactic dynamics, increasing the astrometric precision and sample size by orders of magnitude compared to previous surveys \citep{hipparcos}.

Dynamical mass measurements are typically performed under the assumption of a steady state. However, \emph{Gaia} has made it all the more clear that the Galaxy is host to time-varying dynamical structures. One such structure is the phase-space spiral recently discovered by \citet{2018Natur.561..360A}, seen in the phase-space plane of position and velocity in the direction perpendicular to the Galactic disk, which is present in the solar neighbourhood as well as more distant regions of the Galactic disk \citep{Laporte19,2019ApJ...877L...7W, antoja21}. The phase-space spiral is not in a steady state and thus constitutes a bias to studies that are based on that assumption. However, its presence is not necessarily an obstacle to dynamical mass measurements but it can instead be regarded as an asset because the winding and shape of the phase-space spiral can inform us of the gravitational potential it inhabits.

This work is part of a longer series about weighing the Galactic disk using the phase-space spiral; we have previously published three articles \citep{PaperI,PaperII,PaperIII} which we refer to as \citetalias{PaperI}, \citetalias{PaperII}, and \citetalias{PaperIII}. In these articles, we have tested our method on one-dimensional simulations and applied it to actual \emph{Gaia} data, analysing the immediate solar neighbourhood using the radial velocity sample as well as the distant Galactic disk using the proper motion sample.

This method for weighing the Galactic disk is new, and so far it has only been used in the previous articles of this series. That makes it especially important to test and validate our method on simulations where the answer is known. In this work, we have applied it to a billion particle three-dimensional simulation \citep{2021MNRAS.508.1459H}, as a test of potential sources of bias that could arise in the complex three-dimensional dynamics of a disk galaxy perturbed by an external satellite. Most importantly, we aim to test our model's fundamental assumptions that the phase-space spiral inhabits a gravitational potential that is vertically separable (commonly known as the ``one-dimensional approximation'') and static (neglecting the self-gravity of the spiral perturbation). In addition to the main application of our method, we also ran tests in the presence of strong selection effects, mimicking the incompleteness due to dust extinction and stellar crowding seen in \citetalias{PaperII} and \citetalias{PaperIII}; in this case we included a simple extinction model in our method of inference, similar to the method used in \citetalias{PaperIII}. Furthermore, we tested our method's sensitivity with respect to a biased height of the disk mid-plane.

This article is structured as follows. The details of the simulation are discussed in Sect.~\ref{sec:simulation}. How the data samples are constructed is explained in Sect.~\ref{sec:data}. In Sect.~\ref{sec:model} we discuss our model of inference and in Sect.~\ref{sec:results} we present our results. In Sects.~\ref{sec:discussion} and \ref{sec:conclusion}, we discuss and conclude.

\section{Three-dimensional simulation}\label{sec:simulation}

We use the pure $N$-body simulation which is labelled M1 in \cite{2021MNRAS.508.1459H}, comprised of a Milky Way like host galaxy and a dwarf satellite that merges into it. The initial conditions for the Milky Way like host galaxy was created with the parallelised version of the \texttt{galactics} initial condition generator\footnote{\url{https://github.com/treecode/galactics.parallel}} \citep{KD95}, using the parameters from the Milky Way like model labelled ``MWb'' in \cite{WD05}. This creates a disk which is stable against bar and spiral formation for several billion years (with Toomre parameter $\mathcal{Q}=2.3$). The dwarf galaxy is model L2 of \cite{laporte18}, which is comprised of two Hernquist spheres \citep{H90}. The first represents dark matter and has virial mass $M_{200}=6\times10^{10}~\Msun$, concentration parameter $c_{200}=28$, halo mass $M_h=8\times10^{10}~\Msun$, and scale radius $a_h=8~\kpc$. The second represents the stellar component embedded in the dark halo, and has stellar mass $M_*=6.4\times10^8~\Msun$ and scale radius $a_h=0.85~\kpc$ \citep[see][for a more thorough description]{laporte18}. 

The combined model was evolved for 8.3 Gyr with the GPU based $N$-body tree code \texttt{Bonsai} \citep{Bonsai,Bonsai-242bil}, using a smoothing length of 50 pc and an opening angle $\theta_{\mathrm{o}}=0.4$ radians. In this work we analyse the `present day' snapshot, with $t=6.87$ Gyr, available on \texttt{Flathub}\footnote{\url{https://flathub.flatironinstitute.org/jhunt2021}} as Model M1, snapshot 703. In this time snapshot, the satellite has a phase-space position of  $(x, y, z, u, v, w)=(9.5, -0.4, -6.7, 115.7, 5.9, 311.9)~\kpc$ and $\kmsec$ respectively, in the host galaxy rest frame. It had its most recent pericentre passage 446 Myr before this snapshot, with two disk crossings at 362 and 583 Myr; this is further discussed in Sect.~\ref{sec:results} and shown in Fig.~\ref{fig:hist_t_pert}.

This simulated galaxy is not intended to be a perfect reproduction of the Milky Way and Sagittarius satellite merger, or even the host or satellites themselves. For example, in the simulation's `present day' the satellite is both too massive ($8\times10^9~\Msun$) and too close to the galactic centre compared to the Sgr remnant \citep[e.g.][]{Vasiliev+20}. Instead, it is intended to be a laboratory for studying satellite interaction with an otherwise stable disk, where any non-axisymmetric structure is induced by the satellite interaction.

In terms of phase-space spirals, the simulation has some crucial qualitative differences with respect to the actual Milky Way (as well as the one-dimensional simulations of \citetalias{PaperI}). Although the simulation is very high resolution, the number of stellar particles is still roughly a factor of 300 smaller than the number of stars in the Milky Way, making it more difficult to resolve the phase-space spiral in small spatial volumes. Furthermore, the phase-space spiral seen in the three-dimensional simulation is present at greater vertical energies (i.e. greater vertical velocities and greater heights from the mid-plane). The main reason for this is that the simulation is comprised of particles representing stars and dark matter, but lacks a component of cold gas. In the solar neighbourhood, cold gas has a mid-plane matter density that is roughly equal that of stars (see e.g. \citealt{2015ApJ...814...13M} and \citealt{Schutz:2017tfp}), although it has a significantly smaller scale height (roughly 100~pc). On large spatial scales, the cold gas contribution is not very significant, but it does matter for the formation of the phase-space spiral at smaller vertical energies. For a perturbation to wind into a phase-space spiral, the vertical oscillation period must vary with vertical energy, which requires an anharmonic gravitational potential. The small scale height of the cold gas component makes the vertical gravitational potential anharmonic at smaller heights, such that the winding of the spiral can occur at lower vertical energies.

The main purpose of this work was to test our method when applied to the complex dynamics of a three-dimensional simulation of a disturbed disk galaxy. Specifically, we sought to test our method's fundamental assumptions that the spiral inhabits a static and vertically separable gravitational potential. Due to the properties of the simulation and its phase-space spiral, as discussed above, we would expect any potential bias that could arise from these fundamental assumptions to be amplified: the strength of the perturbation poses a greater challenge to the assumption that the spiral evolves in a static gravitational potential; the spiral's presence at greater heights is more challenging of the assumption of vertical separability; the lower resolution and weaker statistics makes extracting the shape of the spiral less robust. However, as we shall see in Sect.~\ref{sec:results}, our method performed well and produced unbiased results despite these additional difficulties.

\section{Data sample construction}\label{sec:data}

We constructed the data samples analysed in this work from the simulation particles that represent stars. We divided the galaxy's disk plane into a grid in galactocentric longitude and galactocentric radius, with widths of 15 degrees and 500~pc and a radius range of 6--10~kpc. For each grid point, we selected a spatial volume centred on the grid point, extending 300~pc in the radial direction and 600~pc in the azimuthal direction. For each of these respective volumes, we limited ourselves to azimuthal angular momenta within plus or minus ten per cent of that data samples' mean value, similar to the data construction procedure in \citetalias{PaperII}.

We then studied the individual data samples by eye and selected those where a well defined single-armed phase-space spiral was visible. Many disqualified data samples had a phase-space spiral structure which was not very clean, for example with multiple and sometimes fractured arms, likely related to multiple interactions with the orbiting satellite. In a few cases, the phase-space spiral had a clear single arm but with a thwarted shape that could not be reproduced by our fitting algorithm.

\begin{figure}
	\includegraphics[width=1.\columnwidth]{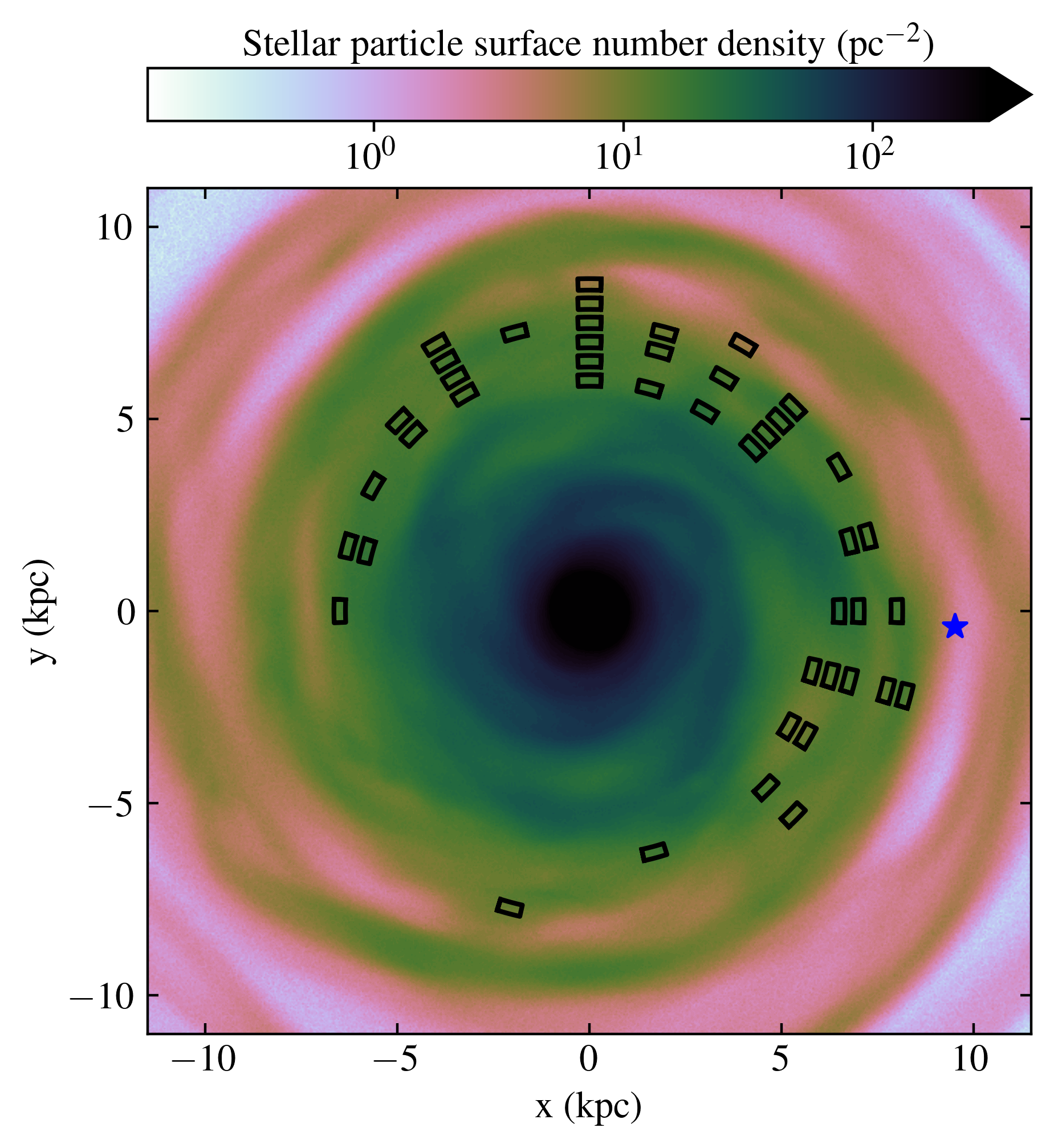}
    \caption{Stellar surface density of the simulation snapshot, with overlaid black squares corresponding to the areas of our data samples. The star marks the location of the perturbing satellite in the $(x,y)$-plane.}
    \label{fig:surf_dens}
\end{figure}

After this screening we had a total number of 44 data samples, whose locations in the disk plane are visible in Fig.~\ref{fig:surf_dens}, overlaid on the stellar surface number density. The data samples' locations have a similar disk surface density; the galaxy's more massive inner regions are less prone to being perturbed due to its stronger self-gravity, while the less massive outer regions are influenced by the satellite over longer time-scales which is less conducive to producing a well defined phase-space spiral. The bottom left quadrant of the disk (negative $x$ and $y$) has almost no useful data samples; this region was most severely affected by the most recent passage of the satellite through the galactic disk.

We also performed additional tests under the influence of severe and spatially dependent selection effects, which can arise mainly due to stellar crowding and dust extinction close to the Galactic mid-plane, similar to what we saw in the \emph{Gaia} data we analysed in \citetalias{PaperIII}. In order to demonstrate the robustness of our method under unknown selection effects, we perform tests where the simulation data was subjected to extinction. In these tests, our method of inference had no information about the precise form of this extinction; rather, the model of inference includes a simple mask model which was fitted in a data driven manner (also used in \citetalias{PaperIII}, see Sect.~\ref{sec:model} for further details).

We randomly generated ten separate extinction functions. They are one-dimensional Gaussian mixture models proportional to
\begin{equation}\label{eq:true_extinction}
    \text{ext.}(z) \propto \sum_{n=1}^8 a_n \dfrac{\exp \bigg[ -\dfrac{(z-z_{\text{ext.},n})^2}{2\sigma_n^2} \bigg]}{\sqrt{2\pi \sigma_n^2}}.
\end{equation}
The eight Gaussian components are composed of two subgroups; for the first group of three Gaussian (and the second group of five Gaussians), $a_n$ were generated from a uniform distribution in range 0.2--0.3 (0.6--1), $z_{\text{ext.},n}$ were generated from a normal distribution centred on the respective data sample's mean value of $z$ and a standard deviation of 200~pc (150~pc), and $\sigma_n$ were generated from a uniform distribution in range 100--200~pc (40--80~pc). After all these parameters have been generated, the extinction function was normalised such that its maximum value corresponds to 50~\%.

\begin{figure}
    \centering
	\includegraphics[width=.8\columnwidth]{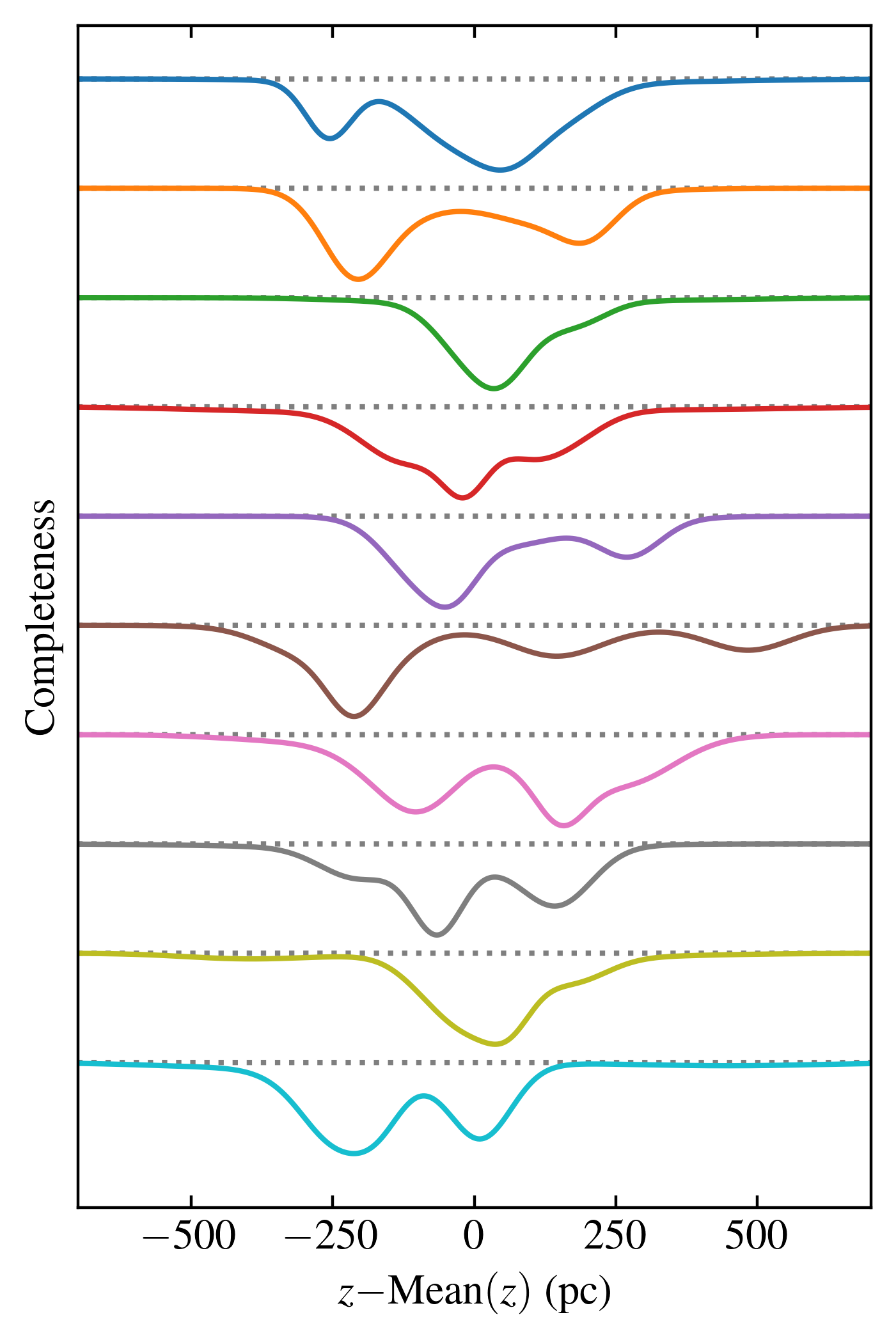}
    \caption{Ten randomly generated extinction functions. They are shown in terms of their completeness as a function of height, where the dotted lines correspond to 100~\% and the bottom of the respective curves correspond to 50~\%.}
    \label{fig:extinction}
\end{figure}

The extinction masks are qualitatively similar to the selection effects observed in \citetalias{PaperIII}, in the sense of being asymmetric and a mix of multiple narrow and broad bands along the $Z$-axis. The selection effects present in the \emph{Gaia} data set vary significantly, depending mainly on distance and Galactic longitude; in \citetalias{PaperIII}, some data samples were not visibly affected while others were rendered completely unusable. The extinction masks constructed in this work correspond to a middle ground, where the extinction masks are by far the dominant feature in the unprocessed data histogram, but the spiral shape can still be robustly extracted in the bulk density fit. The ten randomly generated extinction functions are shown in Fig.~\ref{fig:extinction}.

\section{Model of inference}\label{sec:model}

The model of inference used in this work is generally the same as was used in previous articles in this series. It is most akin to the version used in \citetalias{PaperIII}, although with some minor modifications in scale due to the spiral being present only at greater vertical energies in the three-dimensional simulation (see Sect.~\ref{sec:simulation}). Although the description in this section is complete, we refer back to previous papers for a more extensive explanation of our method.

Some of the details of our inference, such as the area of the $(z,w)$-plane where the spiral is fitted, depend on the standard deviations $\text{Std}(z)$ and $\text{Std}(w)$ for the specific data sample's stellar distribution. For our suite of data samples, these standard deviations are distributed according to $427 \pm 23~\pc$ and $22 \pm 2~\kmsec$.

When applying our method, we reduce the data to a two-dimensional histogram in height and vertical velocity. The phase-space density in our model of inference is a function of these two phase-space coordinates, equal to
\begin{equation}\label{eq:total_density}
    f(z,w\,|\,\popp) = B(z,w\,|\,\popp_\text{bulk}) \times \Big[ 1 + m(z,w)\, S(z-\bar{z},w-\bar{w}\,|\,\popp_\text{spiral}) \Big],
\end{equation}
where $B(z,w\,|\,\popp_\text{bulk})$ is a bulk density distribution, $S(z-\bar{z},w-\bar{w}\,|\,\popp_\text{spiral})$ is a relative spiral density perturbation, and $m(z,w)$ is an inner mask function which quenches the spiral perturbation at low vertical energies. They depend on the free parameters $\popp = \{\popp_\text{bulk},\, \popp_\text{spiral}\}$, which are listed in Table~\ref{tab:model_parameters}.

The bulk density distribution is a Gaussian mixture model equal to
\begin{equation}\label{eq:bulk_density}
    B(z,w\,|\,\popp_\text{bulk}) =
    \sum_{k=1}^{6} a_k \,
    \dfrac{\exp\Bigg[-\dfrac{(z-\bar{z})^2}{2\sigma_{z,k}^2}\Bigg]}{\sqrt{2\pi\sigma_{z,k}^2}} \,
    \dfrac{\exp\Bigg[-\dfrac{(w-\bar{w})^2}{2\sigma_{w,k}^2}\Bigg]}{\sqrt{2\pi\sigma_{w,k}^2}},
\end{equation}
where the Gaussians are constrained to be centred on the same point in the $(z,w)$-plane. Using this centre, we define a translated coordinate system according to
\begin{equation}\label{eq:translated_coords}
    \begin{split}
        Z & \equiv z - \bar{z}, \\ 
        W & \equiv w - \bar{w}.
    \end{split}
\end{equation}

{\renewcommand{\arraystretch}{1.6}
\begin{table}[ht]
	\centering
	\caption{Free parameters of our model. The third sub-group of parameters, written $\popp_{z-\text{mask}}$, is only included for the tests where an extinction function is applied to the data.}
	\label{tab:model_parameters}
    \begin{tabular}{| l | l |}
		\hline
		$\popp_\text{bulk}$  & Bulk phase-space density parameters \\
		\hline
		$a_k$ & Weights of the Gaussian mixture model \\
		$\sigma_{z,k}$, $\sigma_{w,k}$ & Dispersions of Gaussian mixture model \\
		$\bar{z}, \bar{w}$ & Mean height and vertical velocity \\
		\hline
		\hline
		$\popp_\text{spiral}$  & Spiral phase-space density parameters \\
		\hline
		$\rho_{h=\{1,2,3,4\}}$ & Mid-plane matter densities \\
		$t$ & Time since the perturbation was produced \\
		$\tilde{\varphi}_0$ & Initial angle of the perturbation \\
		$\alpha$ & Relative density amplitude of the spiral \\
		\hline
		\hline
		$\popp_{z-\text{mask}}$  & Mask in $z$ (only used for extinction tests) \\
		\hline
		$\hat{a}_l$ & Amplitudes \\
		$\hat{z}_l$ & Means \\
		$\hat{\sigma}_{z,l}$ & Dispersions \\
		\hline
	\end{tabular}
\end{table}}

The spiral density is equal to
\begin{equation}\label{eq:spiral_rel_density}
    S(Z,W\,|\,\popp_\text{spiral}) =
    \alpha \cos\Big[ \varphi(Z,W\,|\, \rho_h)-\tilde{\varphi}(t,E_z \,|\, \rho_h,\tilde{\varphi}_0) \Big].
\end{equation}
In this expression, $E_z = \Phi + W^2/2$ is a vertical energy per mass. It depends on the vertical gravitational potential, which is modelled according to the following functional form,
\begin{equation}\label{eq:phi}
    \Phi(Z\,|\,\rho_h) = \sum_{h=1}^{4}
    4 \pi G \rho_h (2^{h-1}
    \times H)^2 \times \log\Bigg[\cosh\Bigg(\dfrac{Z}{2^{h-1} \times H}\Bigg)\Bigg],
\end{equation}
where the $\rho_h$ parameters are free to vary within the range $[0,0.2]~\Msunppcc$, and $H=3/4 \times \text{Std}(z)$ is proportional to the scale height of the data sample's stellar distribution. The choice of scale heights for the four matter density components differ from previous papers. Here, the scale heights are larger, but due to the lack of a cold gas component in the simulation this function can still faithfully reproduce the shape of the vertical gravitational potential.

The quantity $\varphi$ in Eq.~\eqref{eq:spiral_rel_density} is an angle of vertical oscillation defined like
\begin{equation}\label{eq:angle_of_z}
\begin{split}
& \varphi(Z,W \,|\, \rho_h) = \\
& \begin{cases}
    2 \pi P^{-1}{\displaystyle\int_0^{|Z|}} \dfrac{\de Z'}{\sqrt{2[E_z-\Phi(Z' \,|\, \rho_h)]}} & \text{if}\,Z\geq0\,\text{and}\,W \geq 0, \\
    \pi - 2 \pi P^{-1}{\displaystyle\int_0^{|Z|}} \dfrac{\de Z'}{\sqrt{2[E_z-\Phi(Z' \,|\, \rho_h)]}} & \text{if}\,Z\geq0\,\text{and}\,W<0, \\
    \pi + 2 \pi P^{-1}{\displaystyle\int_0^{|Z|}} \dfrac{\de Z'}{\sqrt{2[E_z-\Phi(Z' \,|\, \rho_h)]}} & \text{if}\,Z<0\,\text{and}\,W<0, \\
    2\pi - 2 \pi P^{-1}{\displaystyle\int_0^{|Z|}} \dfrac{\de Z'}{\sqrt{2[E_z-\Phi(Z' \,|\, \rho_h)]}} & \text{if}\,Z<0\,\text{and}\,W \geq 0.
\end{cases}
\end{split}
\end{equation}
The quantity $\tilde{\varphi}$ is the angle of the spiral as a function of vertical energy, according to
\begin{equation}\label{eq:angle_of_time}
    \tilde{\varphi}(t,E_z \,|\, \rho_h,\tilde{\varphi}_0) = \tilde{\varphi}_0 + 2\pi\frac{t}{P(E_z \,|\, \rho_h)},
\end{equation}
where $P$ is the period of vertical oscillation, equal to
\begin{equation}\label{eq:period}
    P(E_z\,|\, \rho_h) = \oint \frac{\de Z}{W}.
\end{equation}

The inner mask function is equal to
\begin{equation}\label{eq:inner_mask}
    m(z,w) = \sigm \Bigg\{ 10 \Bigg[
    \frac{\big(z-\text{Mean}(z)\big)^2}{\big(3/2 \times \text{Std}(z)\big)^2} + \frac{\big(w-\text{Mean}(w)\big)^2}{\big(3/2 \times \text{Std}(w)\big)^2} -1
    \Bigg] \Bigg\},
\end{equation}
where
\begin{equation}\label{eq:sigmoid}
    \sigm(x) = \frac{1}{1+\exp(-x)}
\end{equation} 
is a sigmoid function. The inner mask function is centred on the respective data samples' mean values of $z$ and $w$, which can differ marginally, typically only by a few parsecs, from the fitted $\bar{z}$ and $\bar{w}$ values. The inner mask function covers a larger area of the $(z,w)$-plane as compared to previous articles in this series (here, denominators include a factor of $3/2$). The reason for this choice is that the phase-space spiral is present at greater vertical energies in the three-dimensional simulation.

In the tests where an extinction function was applied to the data, as described in the end of Sect.~\ref{sec:data}, we also included a simple mask function as part of our model of inference, similar to how selection effects were modelled in \citetalias{PaperIII}. In such a case, the model's phase-space density would instead read
\begin{equation}\label{eq:total_density_with_extinction}
\begin{split}
    f(z,w\,|\,\popp) & = B(z,w\,|\,\popp_\text{bulk}) \times \Xi(z \, | \, \popp_{z\text{-mask}}) \\
    & \times \Big[ 1 + m(z,w)\, S(z,w-\bar{w}\,|\,\popp_\text{spiral}) \Big],
\end{split}
\end{equation}
where
\begin{equation}\label{eq:mask_function}
	\Xi(z \, | \, \popp_{z\text{-mask}}) = 1 - \sum_{l=1}^{6} \hat{a}_l 
	\exp\Bigg[-\dfrac{(z-\hat{z}_l)^2}{2\hat{\sigma}_{z,l}^2}\Bigg].
\end{equation}
The free parameters of the mask function are written with hats, constrained to lie in the ranges $\hat{a}_l \in [0,1]$, $\hat{z}_l-\bar{z} \in [-500,500]~\pc$, and $\hat{\sigma}_{z,l} \in [60,300]~\pc$. This mask function is fitted in a data driven manner and has no information about the precise form of the actual extinction function that was applied to the data prior to inference, apart from prior knowledge that extinction has a purely spatial dependence.

As for previous papers in this series, the minimisation algorithm is run in two separate steps, where the bulk density (without any spiral perturbation) is fitted in a first step, and the spiral density (with a fixed bulk density) is fitted in a second step. The likelihood of the fit is written
\begin{equation}\label{eq:likelihood}
\begin{split}
    \ln\, \mathcal{L}(\data_{i,j}\,|\,\popp) = &
    - \sum_{i,j} M(z_i,w_i) \times \dfrac{[\data_{i,j}-f(z_i,w_i\,|\,\popp)]^2}{2 f(z_i,w_i\,|\,\popp)} \\
    & + \{\text{constant term}\},
\end{split}
\end{equation}
where $\data_{i,j}$ is the two-dimensional data histogram with bin lengths of $0.08 \times \text{Std}(z)$ and $0.08 \times \text{Std}(w)$. The quantity
\begin{equation}\label{eq:outer_mask}
    M(z,w) = 
    \text{sigm} \Bigg\{ -10\, \Bigg[ \frac{\big(z-\text{Mean}(z)\big)^2}{z_\text{lim.}^2} + \frac{\big(w-\text{Mean}(w)\big)^2}{w_\text{lim.}^2} - 1 \Bigg] \Bigg\}
\end{equation}
is an outer mask function, centred on the data sample's mean values of $z$ and $w$, just like for the inner mask function of Eq.~\eqref{eq:inner_mask}. The values for its boundaries are $z_\text{lim.} = 3.5 \times \text{Std}(z)~\pc$ and $w_\text{lim.} = 3.5 \times \text{Std}(w)$ in the first step of our minimisation, and $z_\text{lim.} = 3 \times 300~\pc$ and $w_\text{lim.} = 3 \times \text{Std}(w)$ in the second step. These values are larger in this work compared to previous papers, because the phase-space spiral is present at greater vertical energies in the three-dimensional simulation.

When running the tests where an extinction function was applied to the data, we fitted the mask function as part of the first step of the minimisation algorithm. For these tests, we fixed the value of $\bar{z}$ to the value that was obtained in the standard fit. The reason for doing so is that the parameter $\bar{z}$ cannot be robustly inferred in the presence of these severe extinction, so its value should be informed by some other measurement, which we in these tests assumed to be readily available information.

When running the tests where the height of the disk mid-plane was biased, we used the bulk density that was inferred in our standard fits. We only recomputed the second minimisation step, where the spatial position of the spiral's centre was shifted by some constant $z_\text{bias}$, according to $S(z-\bar{z}+z_\text{bias},w-\bar{w}\,|\,\popp_\text{spiral})$.

\section{Results}\label{sec:results}

In this section, we show the results from having applied our method of inference to the 44 data samples we constructed. We also show the results for the cases when the data was subject to severe selection effects and when the height of the disk mid-plane was biased. When comparing our inferred results with the true potential, that true potential is given by the time snapshot in which the inference was performed, calculated separately for each respective data samples as an average over that data sample's area in the directions parallel to the disk plane (as seen in Fig.~\ref{fig:surf_dens}); hence, this true potential is free from any time averaging or large-scale spatial averaging over for example the azimuthal angle.

Due to the lower resolution of the three-dimensional simulation, we actually have less statistical power in this work as compared to when our model was applied to the Milky Way. For the data samples that we constructed from the three-dimensional simulation (as described in Sect.~\ref{sec:data}), there was on average $1.3 \times 10^4$ stellar particles in the region of the $(z,w)$-plane where the spiral density perturbation was fitted (as defined by the inner and outer mask functions in Sect.~\ref{sec:model}). In comparison, for the real data samples of \citetalias{PaperIII} the corresponding mean number of particles was twice as high, despite covering a significantly smaller spatial volume and being subject to severe selection effects.

\begin{figure*}
	\includegraphics[width=1.\textwidth]{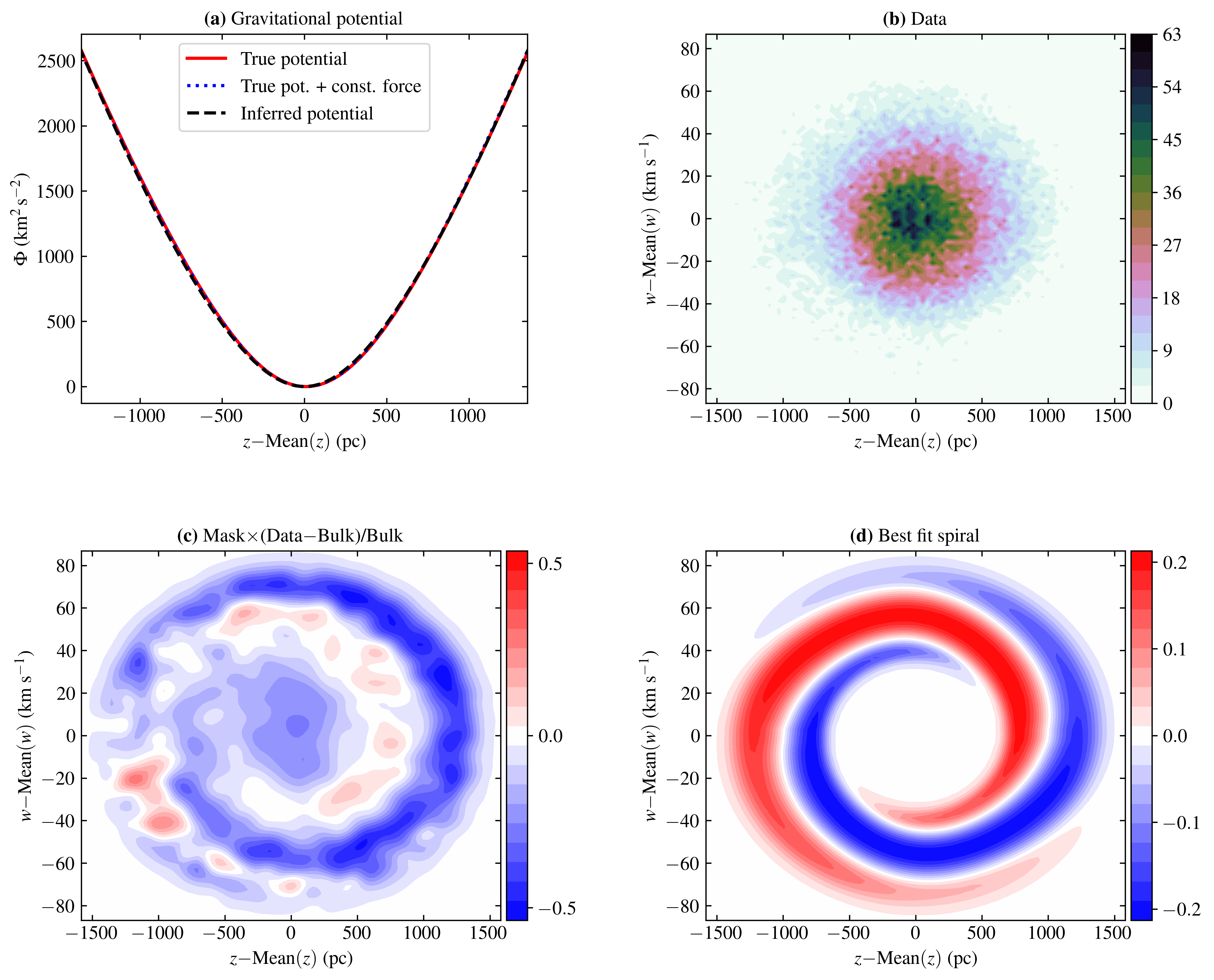}
    \caption{Data and fitted phase-space density of the data sample with $\bar{l} = 165~\deg$ and $\bar{R}=6500~\pc$. The four panels show:
    {\bf (a)} the data histogram;
    {\bf (b)} the fitted bulk density;
    {\bf (c)} the residual phase-space density with respect to the fitted bulk;
    {\bf (d)} the relative phase-space density perturbation of the best fit spiral.}
    \label{fig:spiral_6500_165}
\end{figure*}

In Fig.~\ref{fig:spiral_6500_165}, we show the gravitational potential, data histogram, data residual with respect to the fitted bulk, and the fitted spiral for a representative data sample. The corresponding figures of a few more data samples are found in Appendix~\ref{app:plots}. The fitted spirals seen in panels {\bf (d)} are in good agreements with the shape seen in the data, most clearly in comparison with panels {\bf (c)}. When comparing the inferred gravitational potential with the true potential of the simulation, as seen in panels {\bf (a)}, there is an offset for data samples that are located close to the current position of the satellite. This is seen most clearly in Fig.~\ref{fig:spiral_7500_30}, which is one of the more extreme cases. This external force is not incorporated in our model of inference; in fact, such an external force cannot be inferred from the dynamics in a compact spatial volume alone, as it depends on the assumed boundary conditions. To avoid having to discard the data samples that are in the satellite's vicinity, we added a constant force, corresponding to a constant vertical acceleration field in the data sample's spatial volume, that symmetrises the true potential. Via the Poisson equation, such a constant acceleration field does not affect the underlying matter density field. In this manner, we isolated the gravitational potential that arises from the gravity of the galactic disk. This constant force was chosen such that the true gravitational potential values at $Z=\pm 3\times \text{Std}(z)$ become equal (although the precise choice of height is not significant for our end result). When we do this, we see a much stronger agreement with the inferred gravitational potential for the few data samples in the satellite's vicinity, as exemplified in Fig.~\ref{fig:spiral_7500_30}. For other data samples, this correction is negligible, as can be seen in Fig.~\ref{fig:spiral_6500_165} and other figures in Appendix~\ref{app:plots}. In the remainder of this paper, when we compare the inferred and true gravitational potential, we apply this constant force correction.

\begin{figure}
	\includegraphics[width=1.\columnwidth]{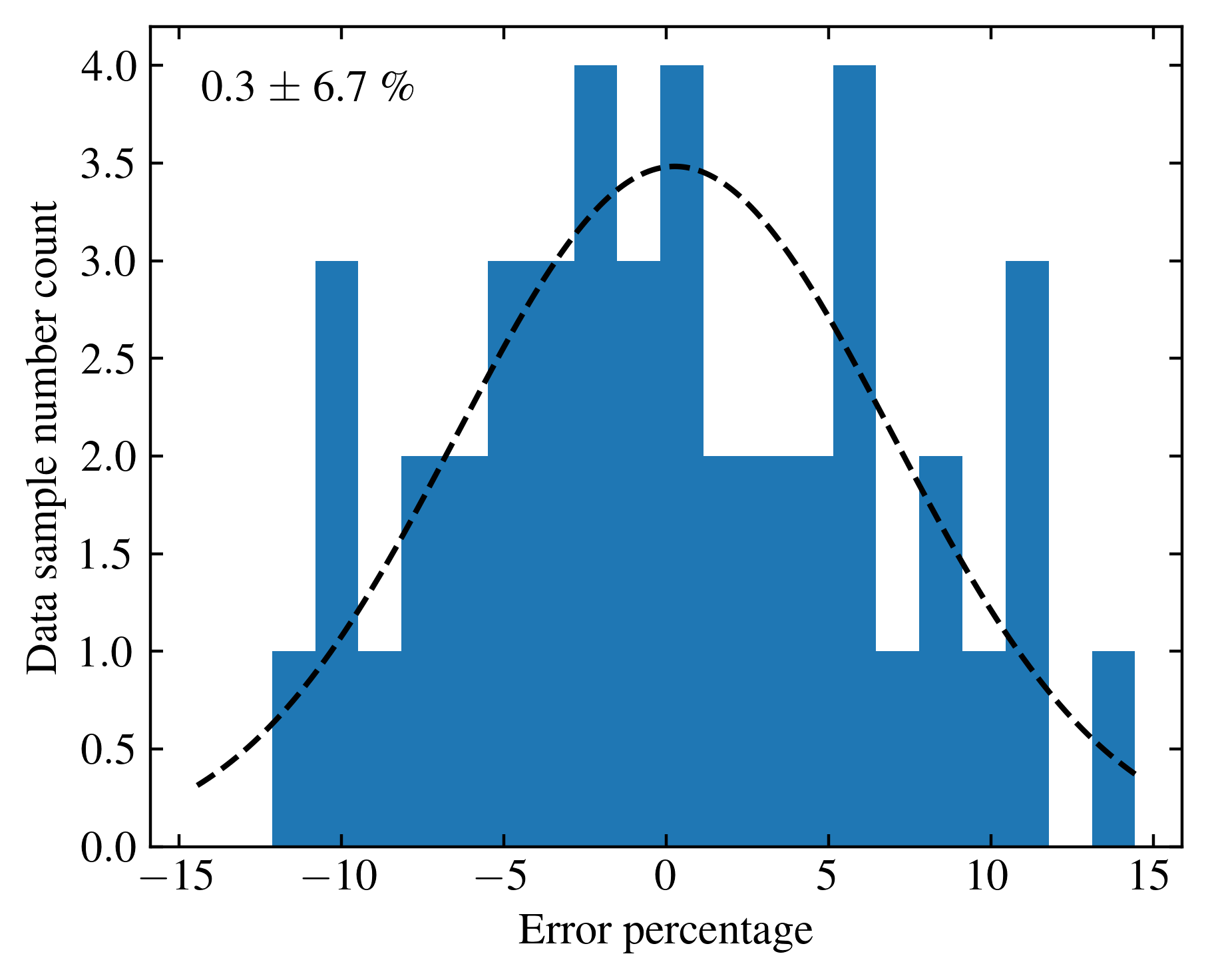}
    \caption{Histogram of relative errors, for the quantity $\Phi(1~\kpc)$. The dashed black line corresponds to a Gaussian distribution whose mean and standard deviation is given by the distribution of relative errors, which are also written in the top left corner.}
    \label{fig:hist_rel_err_1000pc}
\end{figure}

For the inferred gravitational potential, our results are most accurate for the approximate height of $1~\kpc$ (in terms of the potential difference with respect to the mid-plane). We show the relative errors for the inferred quantity $\Phi(1~\kpc)$ for our 44 data samples as a histogram in Fig.~\ref{fig:hist_rel_err_1000pc} and in terms of its distribution in the disk plane in Fig.~\ref{fig:volumes_rel_err_1000pc}. The corresponding plots for $\Phi(800~\pc)$ can be found in Appendix~\ref{app:plots}. For $\Phi(1~\kpc)$, we had a relative accuracy of 7~\% with no indication of any systematic bias, and the errors did not have any strong spatial dependence. For $\Phi(800~\pc)$, we had a relative accuracy of about 8~\%, with a 2~\% bias towards positive errors. This indicates that our method is slightly biased towards gravitational potentials that are too steep close to the disk mid-plane (i.e. biased towards a matter density distribution that is too pinched). We also applied our method using a gravitational potential that was fixed to its true (although momentary) shape in the simulation snapshot. When doing so, the gravitational potential could vary only in terms of its normalisation. In this case we achieved a very similar result, which was non-biased and with an overall accuracy of 7~\%.

\begin{figure}
	\includegraphics[width=1.\columnwidth]{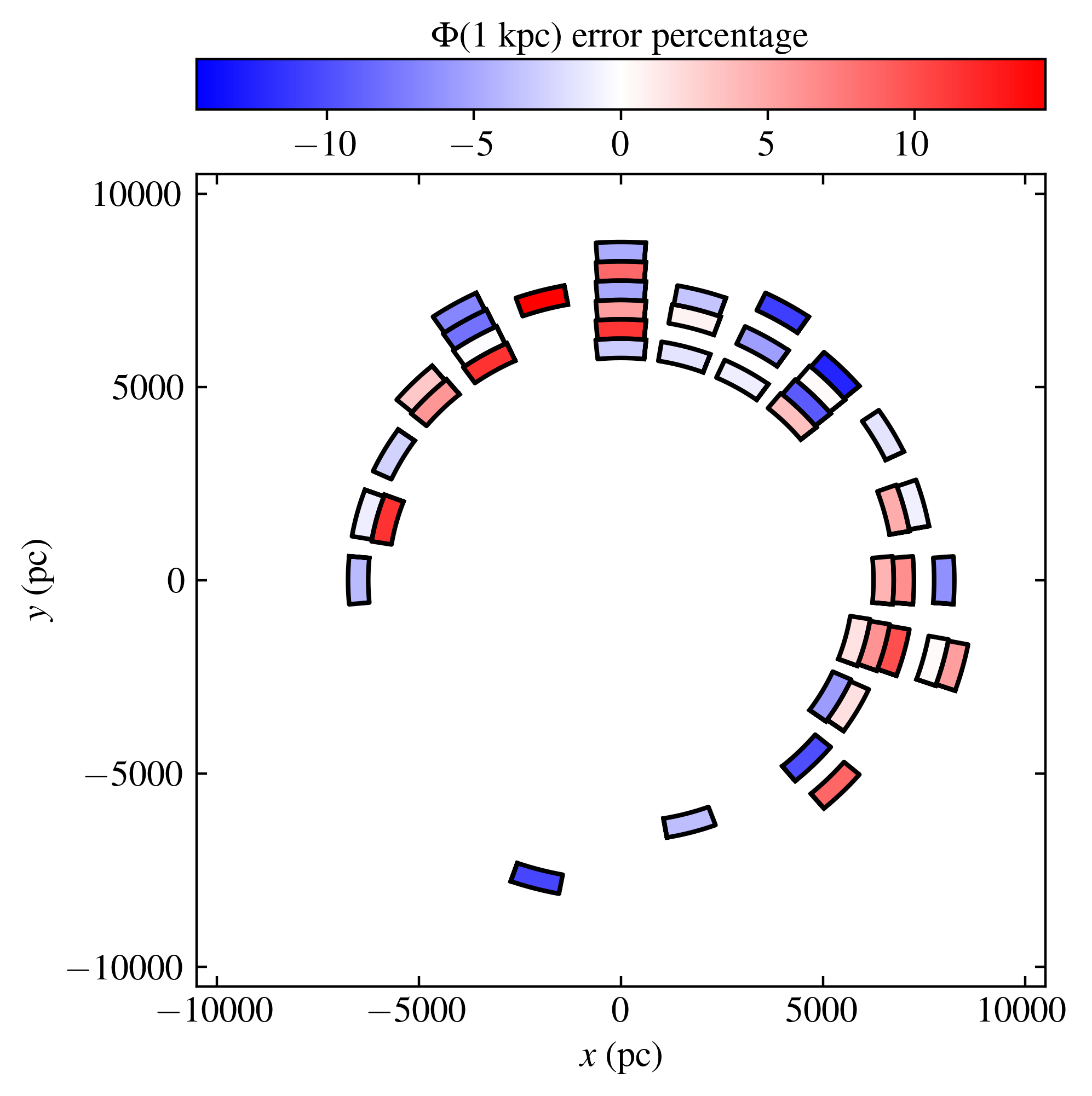}
    \caption{Relative errors of our spiral method for the quantity $\Phi(1~\kpc)$. The areas of the respective data samples are somewhat inflated for better visibility, see Fig.~\ref{fig:surf_dens} for their true size.}
    \label{fig:volumes_rel_err_1000pc}
\end{figure}

\begin{figure}
	\includegraphics[width=1.\columnwidth]{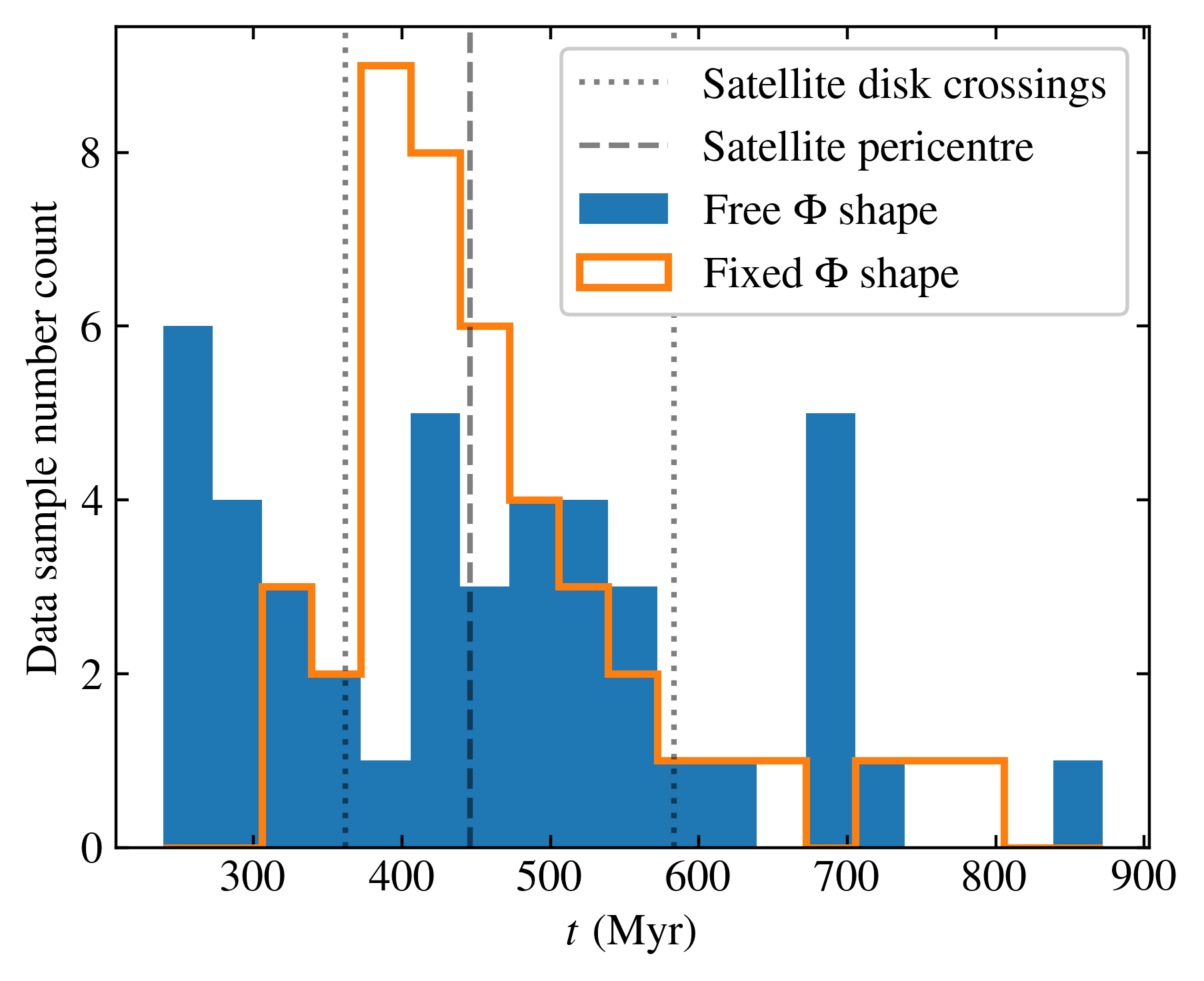}
    \caption{Two histograms of the inferred parameter $t$. The two histograms correspond to fits where the shape of the vertical gravitational potential was either free to vary or fixed to its true shape (free only in terms of its normalisation). The dashed and dotted vertical lines mark the recent close passage of the simulation's satellite.}
    \label{fig:hist_t_pert}
\end{figure}

In Fig.~\ref{fig:hist_t_pert}, we show a histogram of the inferred parameter $t$, which in our model of inference corresponds to the time since the perturbation was produced. In Figs.~\ref{fig:volumes_with_t_pert} and \ref{fig:volumes_with_t_pert_fixed_shape} in Appendix~\ref{app:plots}, we show how these results are distributed in disk. From these figures, it is clear that the strongest outliers are found in the region that is closest to where the perturbing satellite is located. We can compare this with the most recent interactions between the satellite and the inner parts of the Galactic disk; due to its elliptical orbit, the satellite had two recent passages through the disk, at 362 and 583 Myr before the used simulation snapshot, and passed through its pericentre between them at 446 Myr. If we compare these values with the inferred times, they are in rough agreement. As we have discussed in previous papers in this series, see for example \citetalias{PaperI}, the inferred time is highly degenerate with the precise shape of the inferred gravitational potential. If the inference is informed by a strong and correct prior for the shape of the gravitational potential, the time can be inferred with decent accuracy. This is clearly illustrated for the results where the gravitational shape is fixed to its true (although momentary) shape in the simulation snapshot, for which the inferred time is in good agreement with the satellite's most recent close interaction, especially when discounting the few strong outliers which are close at the satellite's current spatial position.

\subsection{Extinction}\label{sec:res_extinction}

\begin{figure}
    \centering
	\includegraphics[width=\columnwidth]{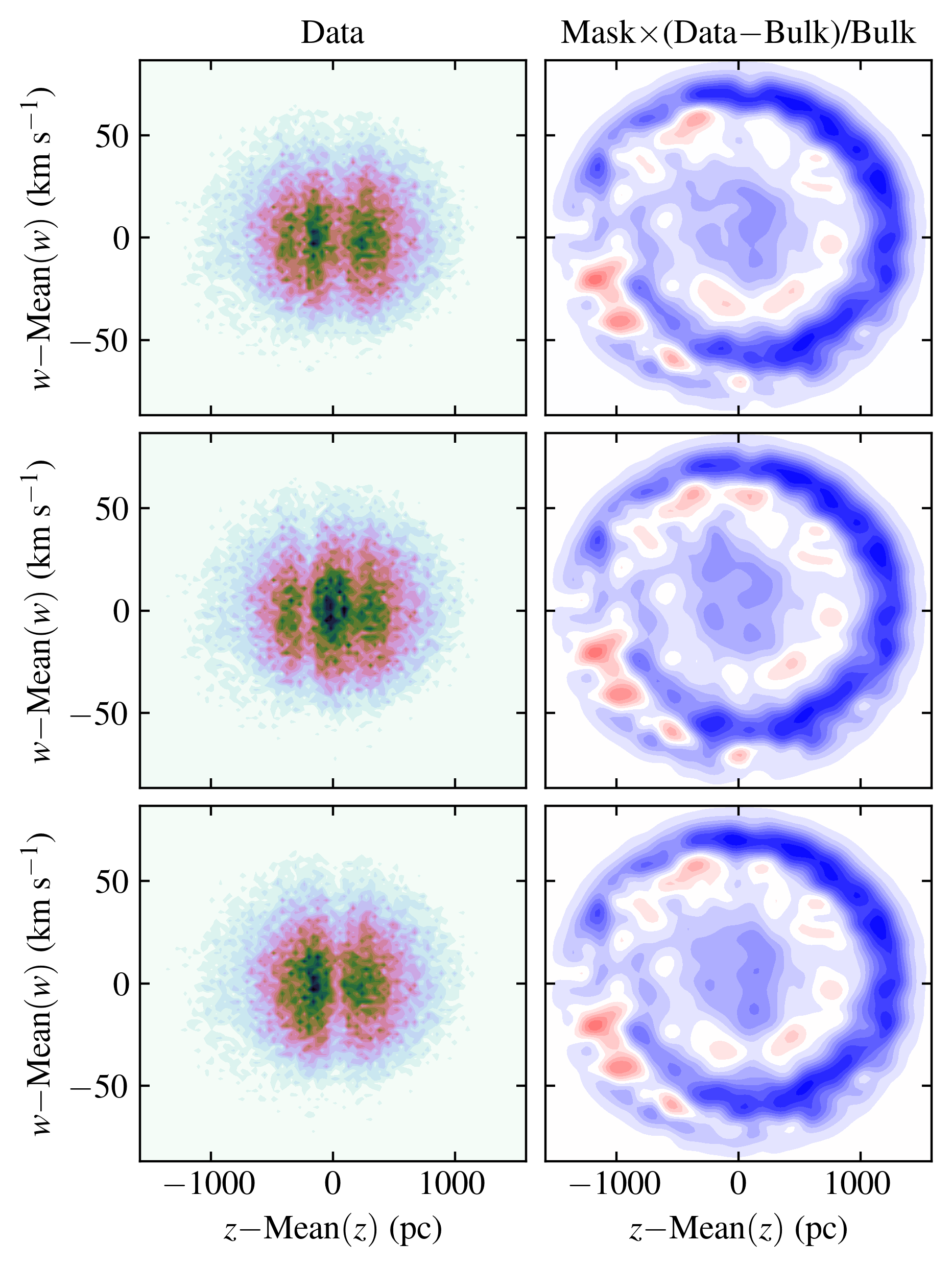}
    \caption{Data histograms affected by extinction functions (which are described in Sect.~\ref{sec:data}. The three examples correspond to extinction functions indices 1--3 (from top to bottom), for the data sample with $\bar{l} = 165~\deg$ and $\bar{R}=6500~\pc$. Left- and right-hand panels correspond to panels {\bf (b)} and {\bf (c)} in Fig.~\ref{fig:spiral_6500_165} (although the ``bulk'' also includes the fitted mask function). The axis ranges are shared between all panels.}
    \label{fig:extinction_examples}
\end{figure}

We applied our ten extinction functions to six of our data samples and then applied our model of inference on these 60 separate cases. In Fig.~\ref{fig:extinction_examples}, we show a few examples of what the data histogram and data residual with respect to the bulk can look like with extinction.

Our method of inference does not have any prior information about the precise form of the extinction function, apart from the knowledge that it depends solely on spatial position. In fact, the mask function which is part of our fitted model, as expressed in Eq.~\eqref{eq:mask_function}, is too simplistic to precisely model the true extinction function of Eq.~\eqref{eq:true_extinction}. In this manner, we test the robustness of our method when incompleteness is severe but unknown and imperfectly modelled. There are degeneracies between the fitted bulk and mask function, which would be detrimental in a traditional dynamical mass measurement based on the assumption of a steady state. However, in our method the inferred gravitational potential is not informed by properties of the bulk density, and depends only on the shape of the extracted spiral. As can be seen in Fig.~\ref{fig:extinction_examples}, the shape of the spiral is robustly inferred despite severe extinction.

\begin{figure}
	\includegraphics[width=1.\columnwidth]{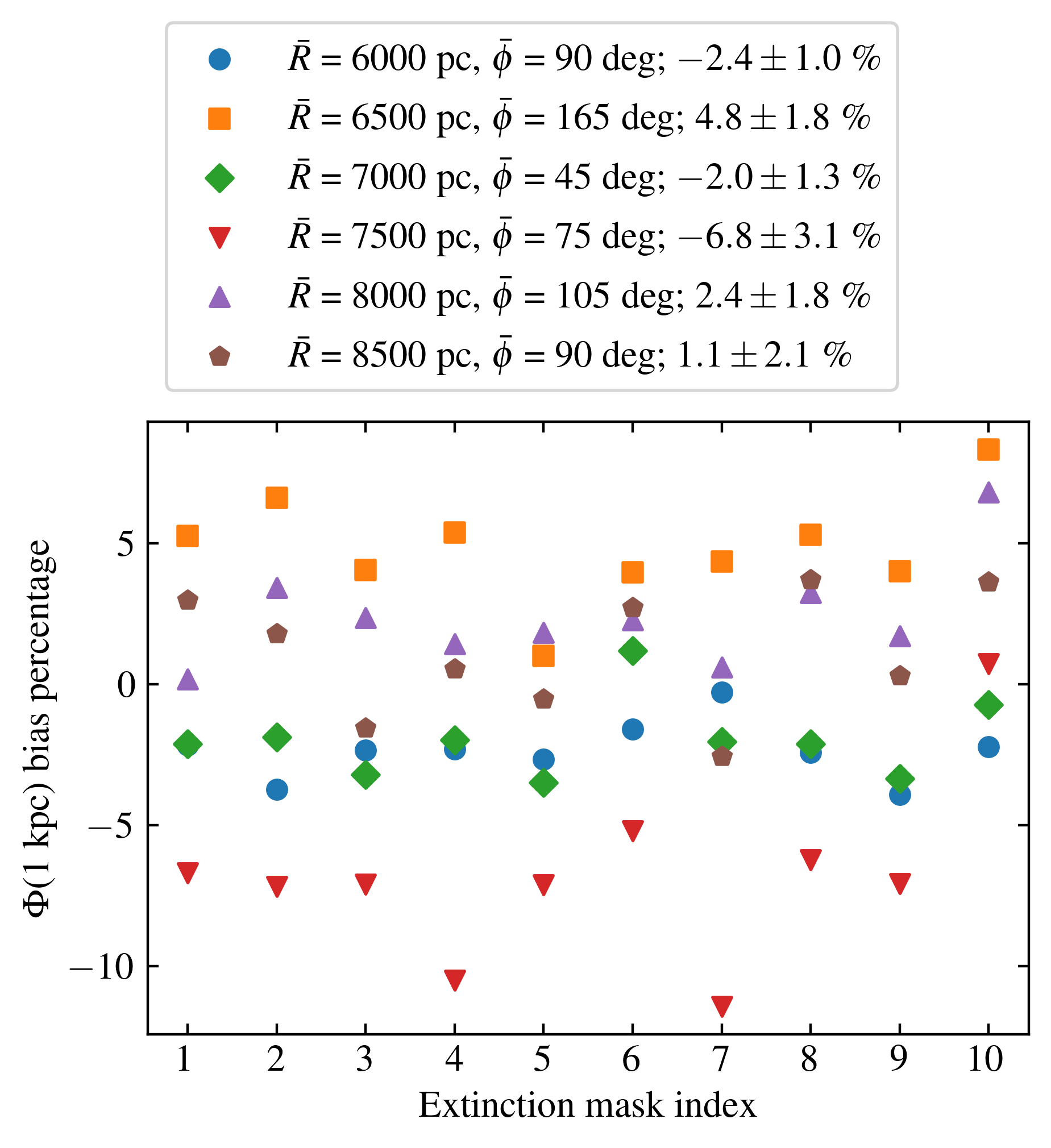}
    \caption{Biases to the inferred quantity $\Phi(1~\kpc)$ that arise when we apply strong selection effects to the data. The bias correspond to the change in $\Phi(1~\kpc)$ with respect to the results with full data completeness. The legend shows the data samples' spatial coordinates as well as the bias mean plus or minus standard deviation.}
    \label{fig:extinction_biases}
\end{figure}

The resulting biases for $\Phi(1~\kpc)$ are shown in Fig.~\ref{fig:extinction_biases}. There is no strong correlation between the biases and the respective extinction functions. Rather, the size and sign of the bias is strongly correlated with the respective data samples. The strongest outlier is the data sample with $\bar{R}=7500~\pc$ and $\bar{\phi}=75~\deg$. The data histogram and fitted phase-space distribution of this data sample can be seen in Fig.~\ref{fig:spiral_7500_75} (for the standard fit without extinction); the fitted spiral does not perfectly reproduce the shape of the spiral seen in the data, which is somewhat skewed along the diagonal of the $(z,w)$-plane. Due to this feature, which our model cannot emulate, it seems that this data sample is especially sensitive to spatially localised selection effects. If we consider the biases of all six data samples and ten extinction masks together, we get a mean plus or minus standard deviation of $-0.5 \pm 4.2~\%$.

In summary, our method performs very well under the influence of severe selection effects and the end result is typically only affected by a few per cent, with no clear trend towards high or low values. This demonstrates that spatially dependent selection effects do not need to be precisely modelled; rather, they need to be modelled just well enough for the shape of the phase-space spiral to be extracted.

\subsection{Biased height of the disk mid-plane}\label{sec:res_zbias}

We tested our method's sensitivity to a biased mid-plane height by imposing a shift in the phase-space spiral's centre, by $\pm 50~\pc$ or $\pm 100~\pc$. We did so for six data samples (the same six that were used in our extinction tests in Sect.~\ref{sec:res_extinction}), giving a total of 24 separate cases.

\begin{figure}
	\includegraphics[width=1.\columnwidth]{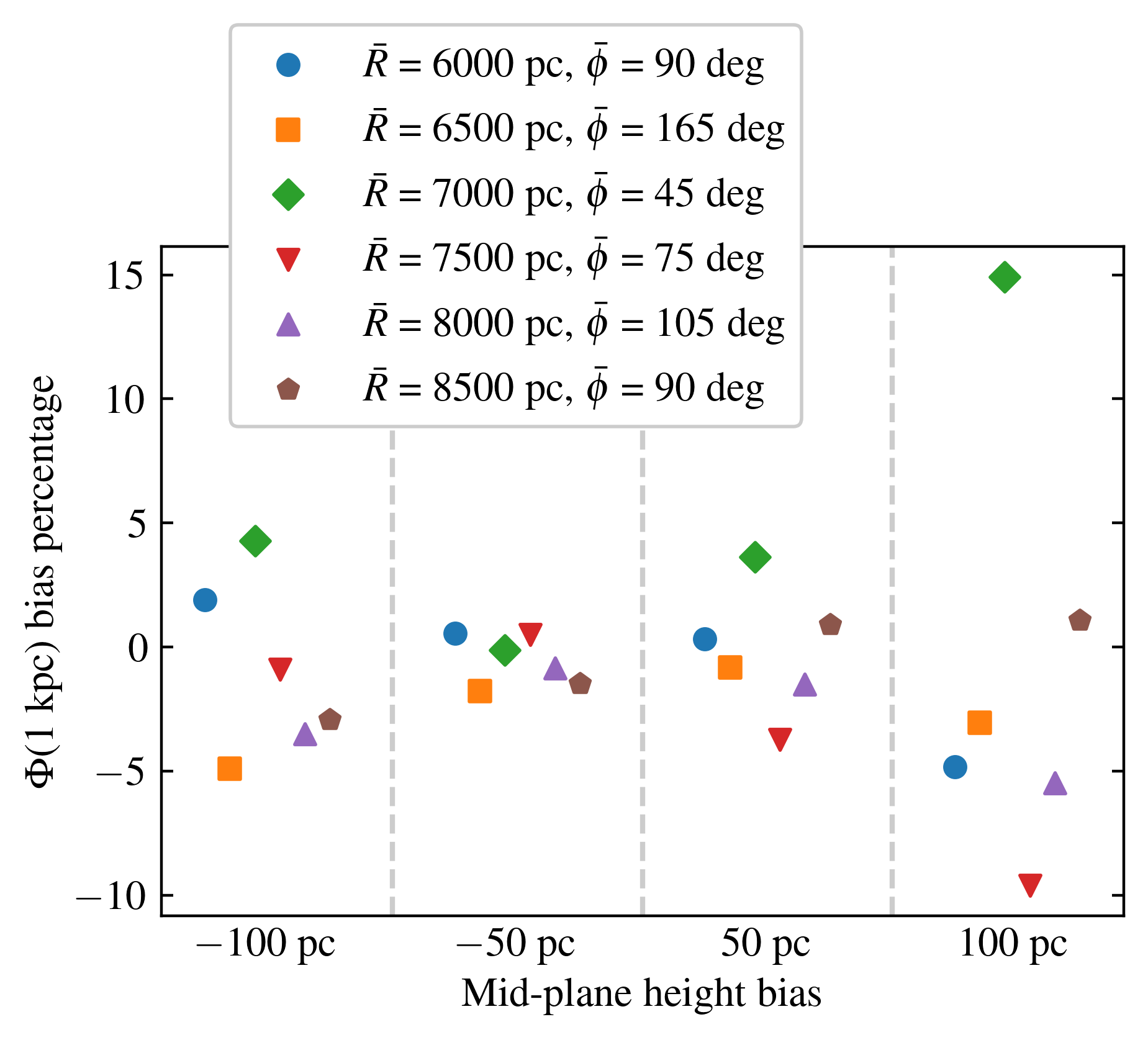}
    \caption{Biases to the inferred quantity $\Phi(1~\kpc)$ that arise when the height of the disk mid-plane is biased.}
    \label{fig:mid_z_biases}
\end{figure}

The resulting bias for the inferred quantity $\Phi(1~\kpc)$, with respect to the case with no added bias in the height of the disk mid-plane, can be seen in Fig.~\ref{fig:mid_z_biases}. For the height bias of $\pm 50~\pc$ ($\pm 100~\pc$), $\Phi(1~\kpc)$ is affected only minimally, with a mean plus or minus standard deviation of $-0.4 \pm 1.7~\%$ ($-1.1 \pm 6.0~\%$). These values also differ between positive and negative height biases, probably related to the fact that the phase-space spiral itself is an asymmetric structure. The data histograms and fitted spirals of the two strongest outliers ($\bar{R}=7000~\pc$ and $\bar{\phi}=45~\deg$, and $\bar{R}=7500~\pc$ and $\bar{\phi}=75~\deg$) can be seen in Figs.~\ref{fig:spiral_7000_45} and \ref{fig:spiral_7500_75}. Both of these data samples have phase-space spirals that are somewhat skewed along the diagonal of the $(z,w)$-plane, potentially related to their spatial proximity to the perturbing satellite. This feature is not reproduced by our model of inference, which is probably related to why these two data samples are especially sensitive to systematic biases.

Our inferred results are stable, despite the large height biases. However, the scale of these biases does not translate perfectly to the case of the actual Milky Way. As is discussed in detail in Sect.~\ref{sec:simulation}, the simulation differs from the Milky Way in some significant regards, mainly due to the simulation's lower resolution and lack of a cold gas disk component, which makes the phase-space spiral present only at higher vertical energies (i.e. greater heights and vertical velocities). For the simulation studied in this work, the most accurate results for the vertical gravitational potential are found at a height of roughly 1~kpc, while for our one-dimensional simulations \citepalias{PaperI} and the Milky Way \citepalias{PaperII,PaperIII}, the most accurate results are found at roughly half that height. Thus the mid-plane height bias applied to the simulation in this work should probably be rescaled by a factor of roughly one half when considering the Milky Way.

\section{Discussion}\label{sec:discussion}

We have tested our new method for weighing the Galactic disk using phase-space spirals on a three-dimensional simulation. As discussed in Sect.~\ref{sec:simulation}, the properties of the simulation differ in some significant regards from the Milky Way, mainly in terms of a stronger external perturbation, lower resolution, less statistics, and a spiral that inhabits greater vertical energies. These simulation properties should be more challenging for our method, especially in terms of its fundamental assumptions of a vertically separable and static gravitational potential, likely amplifying statistical uncertainties and any potential biases in our inferred results. Despite these difficulties, we obtain unbiased results for the vertical gravitational potential with a relative accuracy of 7~\%.

Our fits produced an inferred parameter $t$, corresponding to the time since the perturbation that gives rise to the phase-space spiral was produced. This time can be compared to the satellite most recent pericentre passage (446 Myr before the simulation snapshot that we used) and its two most recent crossings through the host galaxy's disk (362 and 583 Myr). The inferred times agreed reasonably well with the range in time of the satellite's passages, especially for the case where the gravitational potentials of our model were fixed to their true shapes in the simulation snapshot. This illustrates how the inferred time is highly degenerate with the precise shape of the gravitational potential, which is not very robustly inferred relative to for example $\Phi(1~\kpc)$. In our analysis, the gravitational potential has been free to vary in shape; however, going forward we might be better off by fixing the shape of the gravitational potential (or at least applying a more constrained prior), since not much additional information is gathered by letting it be free to vary.

We tested our method under severe spatially dependent selection effects, which produced biases of a few per cent. The extinction masks that we applied to the data were randomly generated in order to mimic the asymmetric incompleteness that arises due to stellar crowding and dust extinction close to the Galactic mid-plane. In these tests, we included a simple extinction function as part of our model of inference \citepalias[as in][]{PaperIII}, which had no prior knowledge about the true form of extinction mask. The selection effects are modelled jointly by the extinction function and the bulk density distribution, where the latter absorbs any incompleteness component that is smooth and reasonably symmetric across the galactic mid-plane. Therefore, the properties of the bulk is highly degenerate with selection, such that probably neither of them are accurately inferred. This is not necessarily a problem but rather a strength of our method, because the bulk is not required to fulfil the collisionless Boltzmann equation, and the gravitational potential is extracted only from the shape of the phase-space spiral. This is a major qualitative difference with respect to traditional dynamical mass measurements that are based on the assumption of equilibrium, where the end result can only be as accurate as the modelling of the selection function. For our method using spirals, we should be able to apply it to great distances and depths and make strong cuts in for example data quality without having to carefully understand or model the relevant selection effects.

We also tested our method under a biased height of the disk mid-plane, by shifting the centre of the fitted spiral along the $z$-axis by $\pm 50~\pc$ and $\pm 100~\pc$. In terms of $\Phi(1~\kpc)$, this produced biases that were mostly contained within a few per cent, with only a few more significant outliers. Overall, our method seems fairly robust with respect to a misunderstood height of the disk mid-plane, at least if the shift is smaller than one tenth of the height where the gravitational potential is measured. As we have seen, selection effects are not a severe obstacle to extracting the phase-space spiral, but could still be detrimental to determining the height of the disk mid-plane. For this reason, this height is better informed by a separate dedicated analysis, which can map the warping of the Galactic disk over a larger spatial area than is covered by our respective data samples.

In terms of the previous articles in this series, this work provides further support of our method and of our results based on \emph{Gaia} data. In the near future, we plan to revisit and expand on the Milky Way analyses that we performed in \citetalias{PaperII} and \citetalias{PaperIII}. In the first half year of 2022, \emph{Gaia} will have its full third data release, which will contain a total of 33 million radial velocity measurements\footnote{\url{https://www.cosmos.esa.int/web/gaia/dr3}}, as compared to the current 7.6 million. Furthermore, complementary distance information has been produced using photo-astrometric measurements by for example \texttt{StarHorse} \citep{2021arXiv211101860A}, claiming a distance precision of roughly 3~\% even for objects with poor parallax measurements; this will allow us to construct data samples with softer cuts in parallax precision, which otherwise induces significant selection effects. With this additional information, we will be able to reach significantly greater depths and distances in the Milky Way's stellar disk, compared to the roughly 3~kpc distance reached in \citetalias{PaperIII}. Thanks to the robustness of our method, we should be able to weigh the Galactic disk at distances that would otherwise not be reachable with for example Jeans analysis.

\section{Conclusion}\label{sec:conclusion}

We have applied our method for weighing the Galactic disk using phase-space spirals to a billion particle three-dimensional simulation of a Milky Way like host galaxy and a merging dwarf satellite. We constructed 44 separate data samples in different locations of the host galaxy's disk, where well defined spirals could be identified by eye. Despite having less statistics and a stronger satellite perturbation than is the case for the actual Milky Way, we obtained non-biased results with a relative accuracy of 7~\%. This validates the most important assumptions of our model of inference, which is that the spiral inhabits a vertically separable and static gravitational potential.

We also tested our method under severe and unknown spatially dependent selection effects, mimicking the incompleteness that can arise from stellar crowding and dust extinction. We obtained accurate results, demonstrating that our method is robust to such effects as long as the shape of the phase-space spiral is accurately extracted. This is a significant qualitative difference with respect to traditional techniques that are based on the assumption of a steady state (e.g. Jeans analysis), for which an accurate modelling of selection effects is crucial. Hence, our method is not only complementary in the sense that it extracts information from a time-varying dynamical structure and therefore subject to different systematic biases, it also has special merit in that it can be applied to distant regions of the Galactic disk where selection effects are difficult to model.

We will apply our method to future \emph{Gaia} data releases, most imminently its full third data release, also supplemented with photo-astrometric measurements (e.g. \texttt{StarHorse}). With greater data depth and precision, we expect to make precise and localised mass measurements of the Galactic disk at even greater distances, further away than what can be reached with other methods.

\begin{acknowledgements}
AW acknowledges support from the Carlsberg Foundation via a Semper Ardens grant (CF15-0384). JH is supported by a Flatiron Research Fellowship at the Flatiron institute, which is supported by the Simons Foundation.
CL acknowledges funding from the European Research Council (ERC) under the European Union's Horizon 2020 research and innovation programme (grant agreement
No. 852839).
GM acknowledges funding from the Agence Nationale de la Recherche (ANR project ANR-18-CE31-0006 and ANR-19-CE31-0017) and from the European Research Council (ERC) under the European Union's Horizon 2020 research and innovation programme (grant agreement No. 834148).
This work made use of an HPC facility funded by a grant from VILLUM FONDEN (projectnumber 16599).

%This work has made use of data from the European Space Agency (ESA) mission \emph{Gaia} (\url{https://www.cosmos.esa.int/gaia}), processed by the \emph{Gaia} Data Processing and Analysis Consortium (DPAC, \url{https://www.cosmos.esa.int/web/gaia/dpac/consortium}). Funding for the DPAC has been provided by national institutions, in particular the institutions participating in the \emph{Gaia} Multilateral Agreement.

This research utilised the following open-source Python packages: \textsc{Matplotlib} \citep{matplotlib}, \textsc{NumPy} \citep{numpy}, \textsc{SciPy} \citep{scipy}, \textsc{Pandas} \citep{pandas}, \textsc{TensorFlow} \citep{tensorflow2015-whitepaper}.
\end{acknowledgements}

%-------------------------------------------------------------------
% \begin{thebibliography}{}

%   \bibitem[1966]{baker} Baker, N. 1966,
%       in Stellar Evolution,
%       ed.\ R. F. Stein,\& A. G. W. Cameron
%       (Plenum, New York) 333

% \end{thebibliography}

\nocite{PaperOne}
\nocite{PaperTwo}
\nocite{PaperThree}
% for the bibliography, at the end
\bibliographystyle{aa} % style aa.bst
\bibliography{thisbib} % your references Yourfile.bib

\begin{appendix} %First online appendix

\section{Supplementary plots}\label{app:plots}

\begin{figure}
	\includegraphics[width=1.\columnwidth]{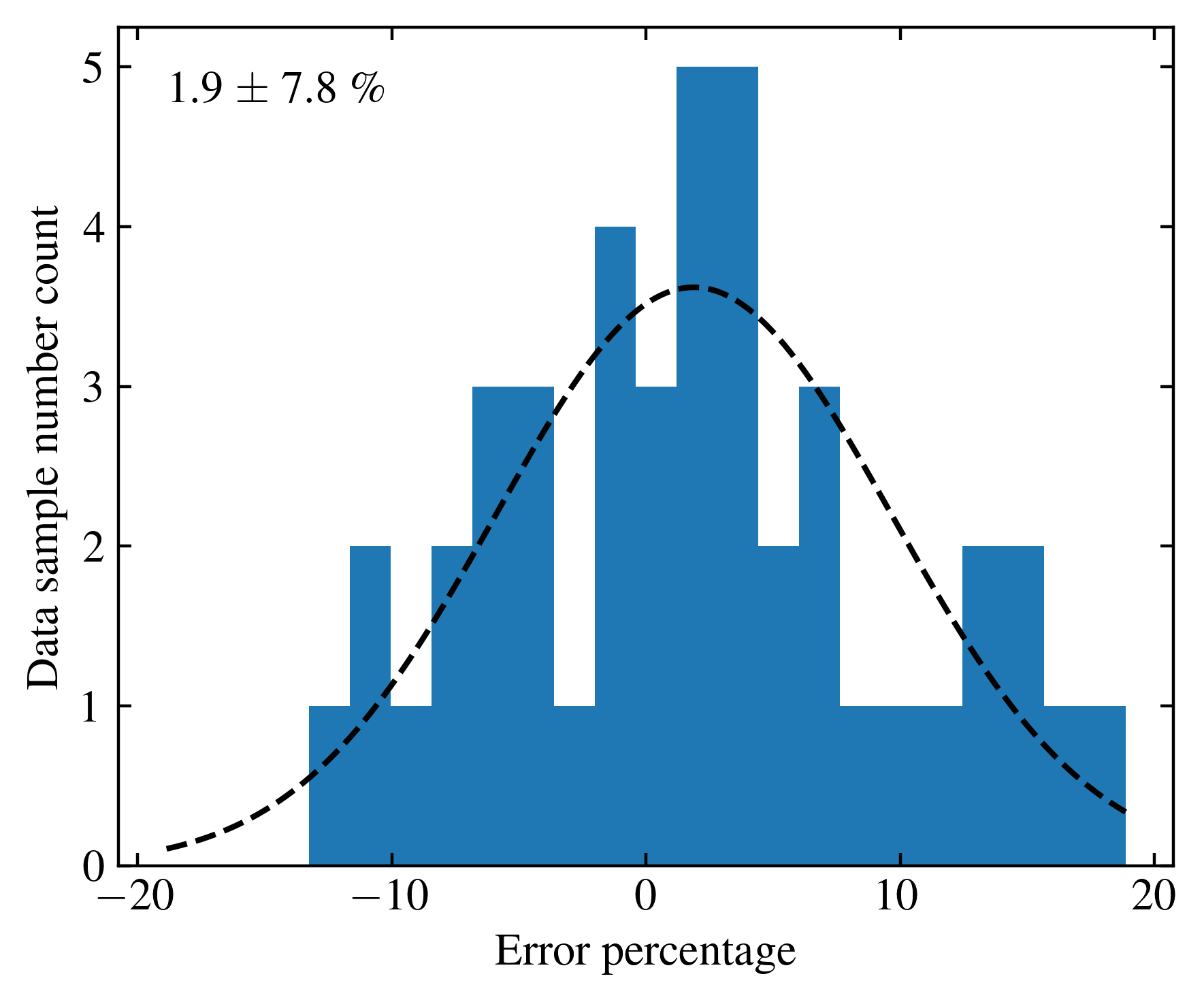}
    \caption{Histogram of relative errors, for the quantity $\Phi(800~\pc)$. The dashed black line corresponds to a Gaussian distribution whose mean and standard deviation is given by the distribution of relative errors, which are also written in the top left corner.}
    \label{fig:hist_rel_err_800pc}
\end{figure}

\begin{figure*}
	\includegraphics[width=1.\textwidth]{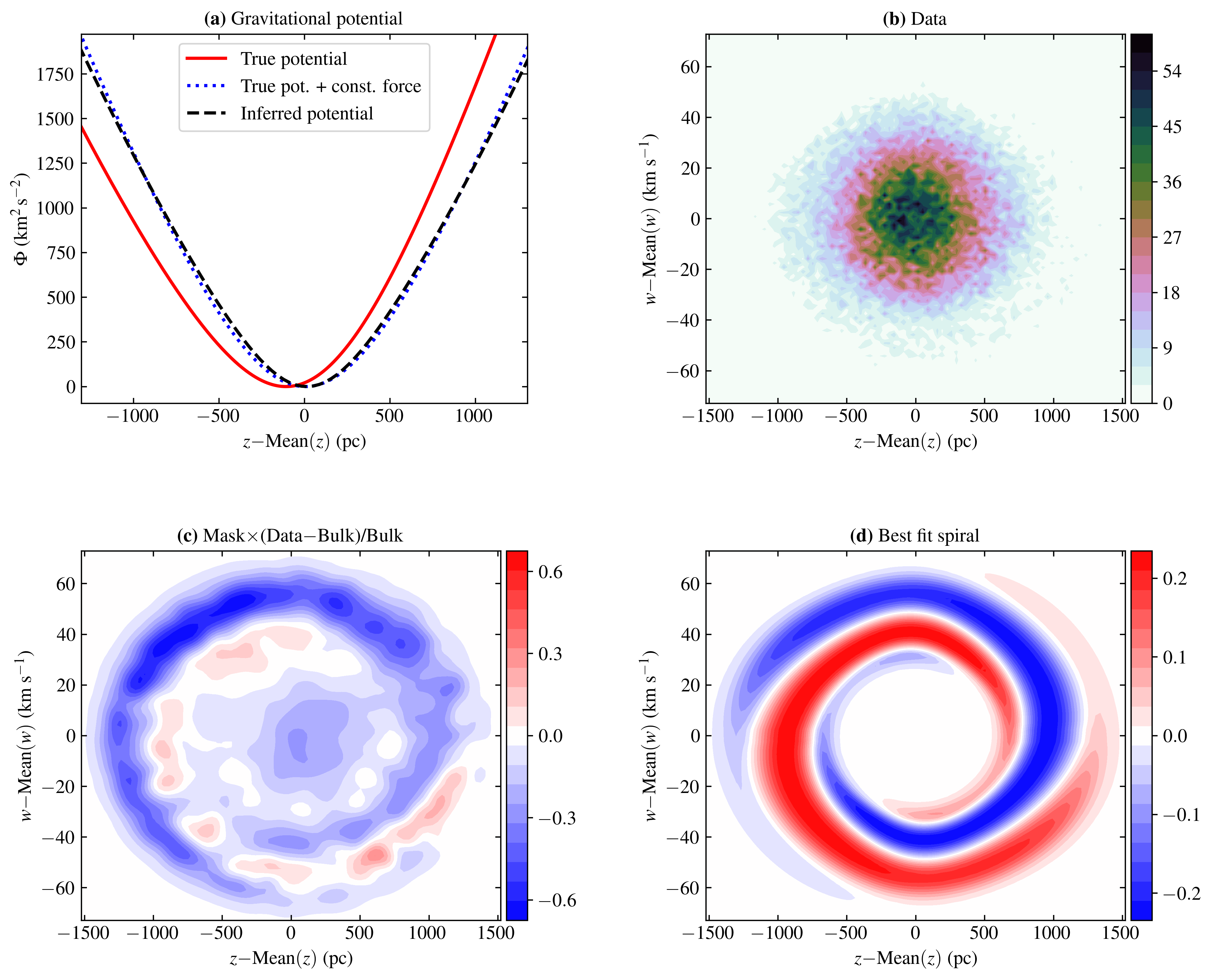}
    \caption{Same as for Fig.~\ref{fig:spiral_6500_165}, but for the data sample with $\bar{l} = 30~\deg$ and $\bar{R}=7500~\pc$.}
    \label{fig:spiral_7500_30}
\end{figure*}

\begin{figure*}
	\includegraphics[width=1.\textwidth]{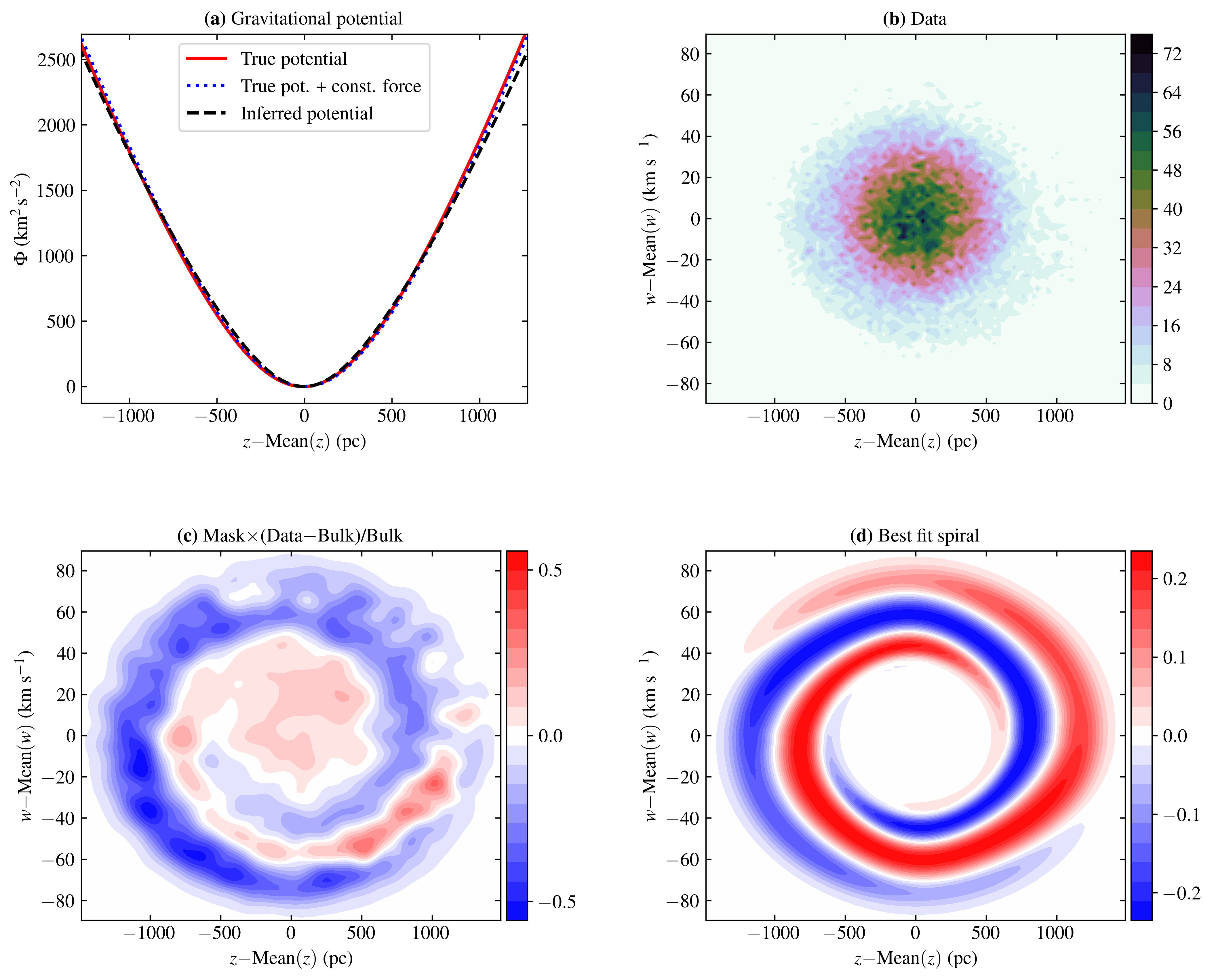}
    \caption{Same as for Fig.~\ref{fig:spiral_6500_165}, but for the data sample with $\bar{l} = 90~\deg$ and $\bar{R}=6000~\pc$.}
    \label{fig:spiral_6000_90}
\end{figure*}

\begin{figure*}
	\includegraphics[width=1.\textwidth]{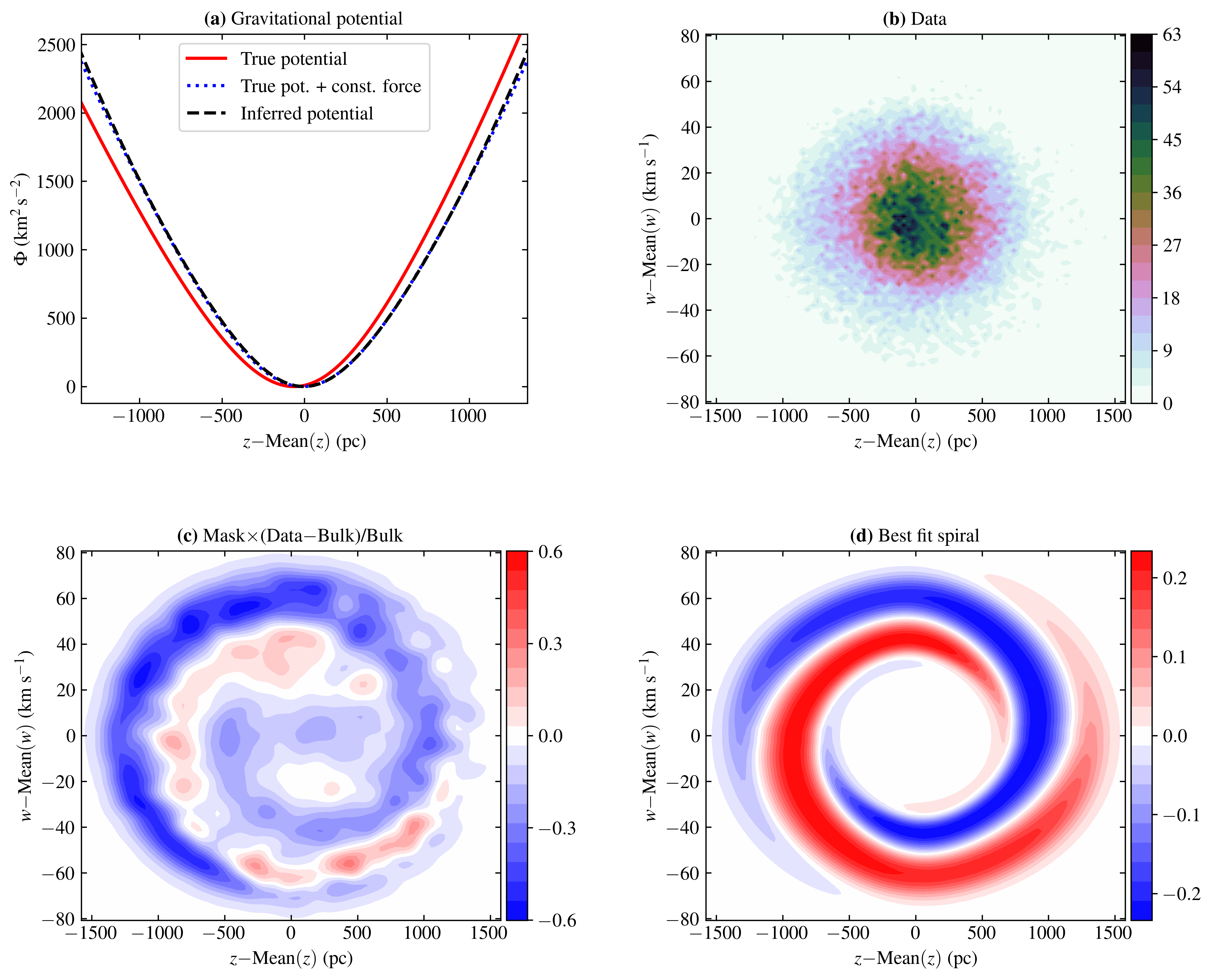}
    \caption{Same as for Fig.~\ref{fig:spiral_6500_165}, but for the data sample with $\bar{l} = 45~\deg$ and $\bar{R}=7000~\pc$.}
    \label{fig:spiral_7000_45}
\end{figure*}

\begin{figure*}
	\includegraphics[width=1.\textwidth]{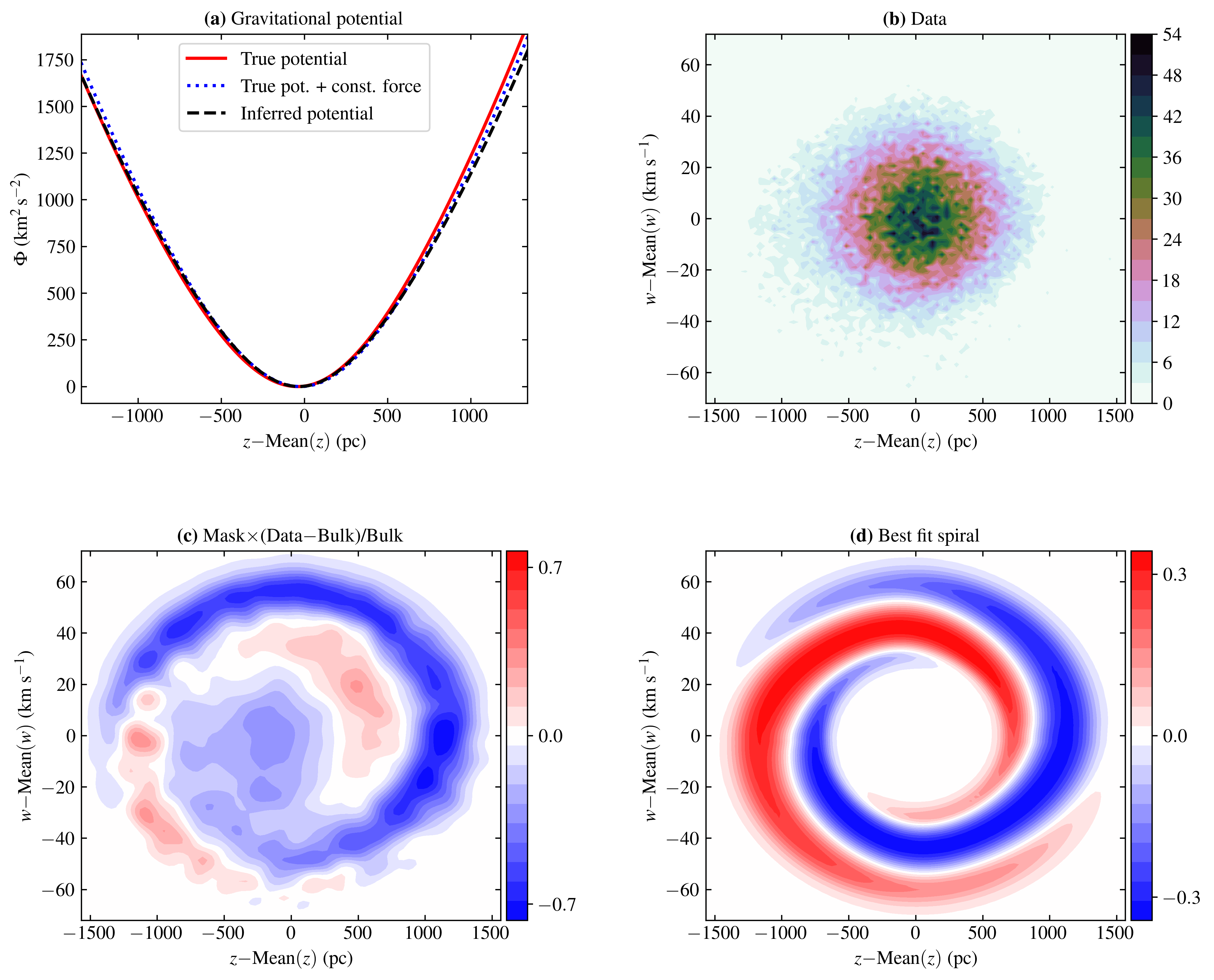}
    \caption{Same as for Fig.~\ref{fig:spiral_6500_165}, but for the data sample with $\bar{l} = 75~\deg$ and $\bar{R}=7500~\pc$.}
    \label{fig:spiral_7500_75}
\end{figure*}

\begin{figure*}
	\includegraphics[width=1.\textwidth]{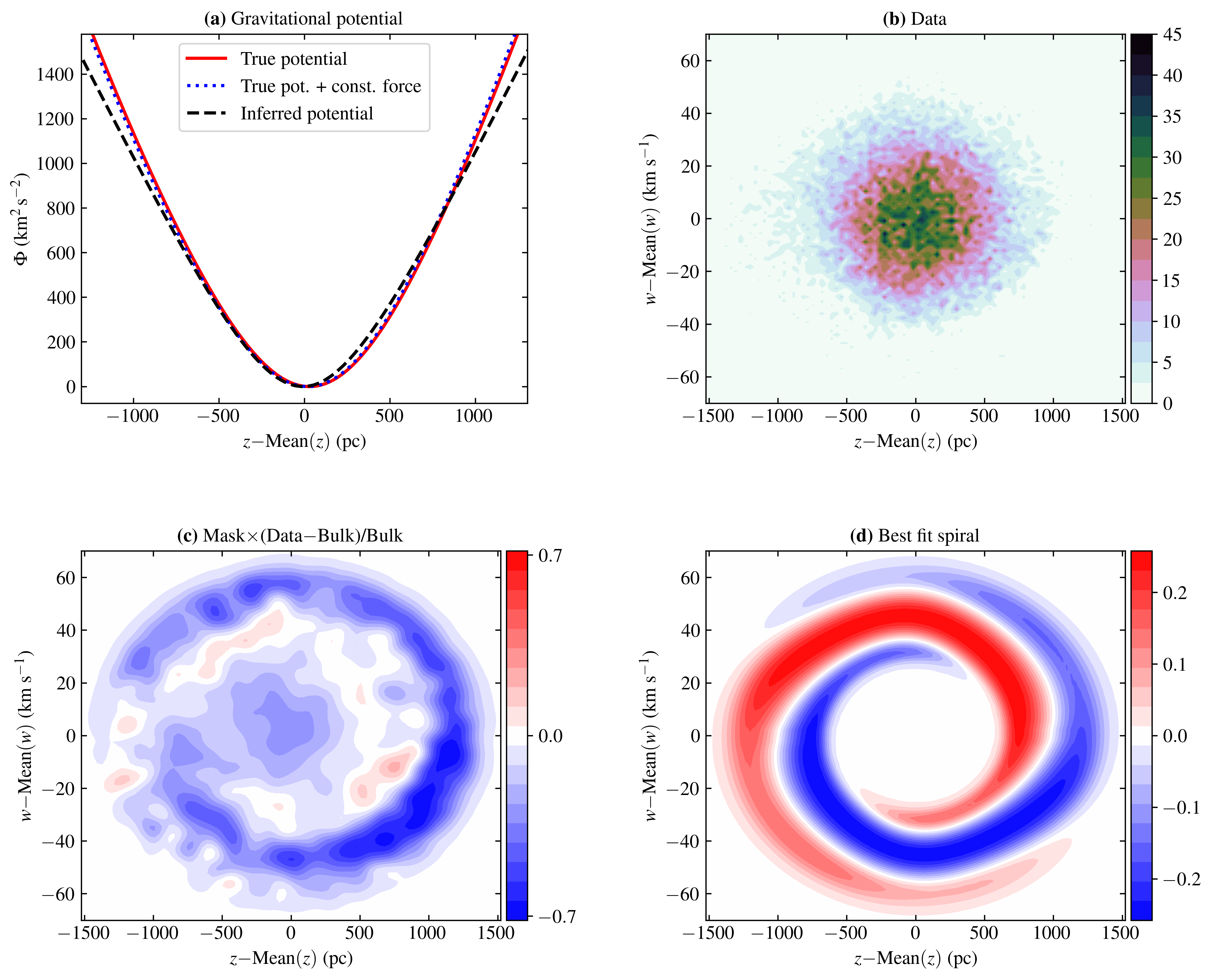}
    \caption{Same as for Fig.~\ref{fig:spiral_8000_105}, but for the data sample with $\bar{l} = 105~\deg$ and $\bar{R}=8000~\pc$.}
    \label{fig:spiral_8000_105}
\end{figure*}

\begin{figure*}
	\includegraphics[width=1.\textwidth]{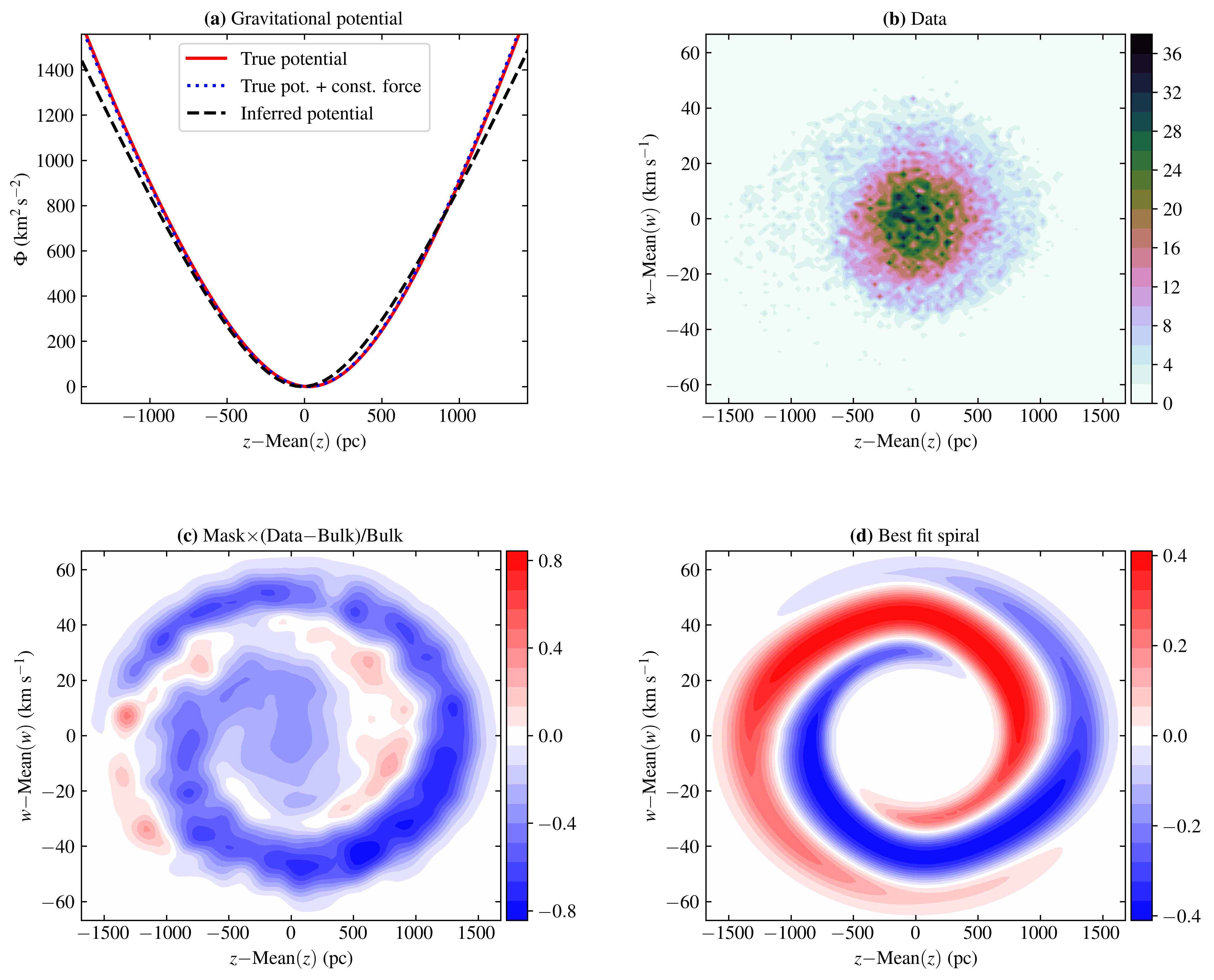}
    \caption{Same as for Fig.~\ref{fig:spiral_6500_165}, but for the data sample with $\bar{l} = 90~\deg$ and $\bar{R}=8500~\pc$.}
    \label{fig:spiral_8500_90}
\end{figure*}

\begin{figure}
	\includegraphics[width=1.\columnwidth]{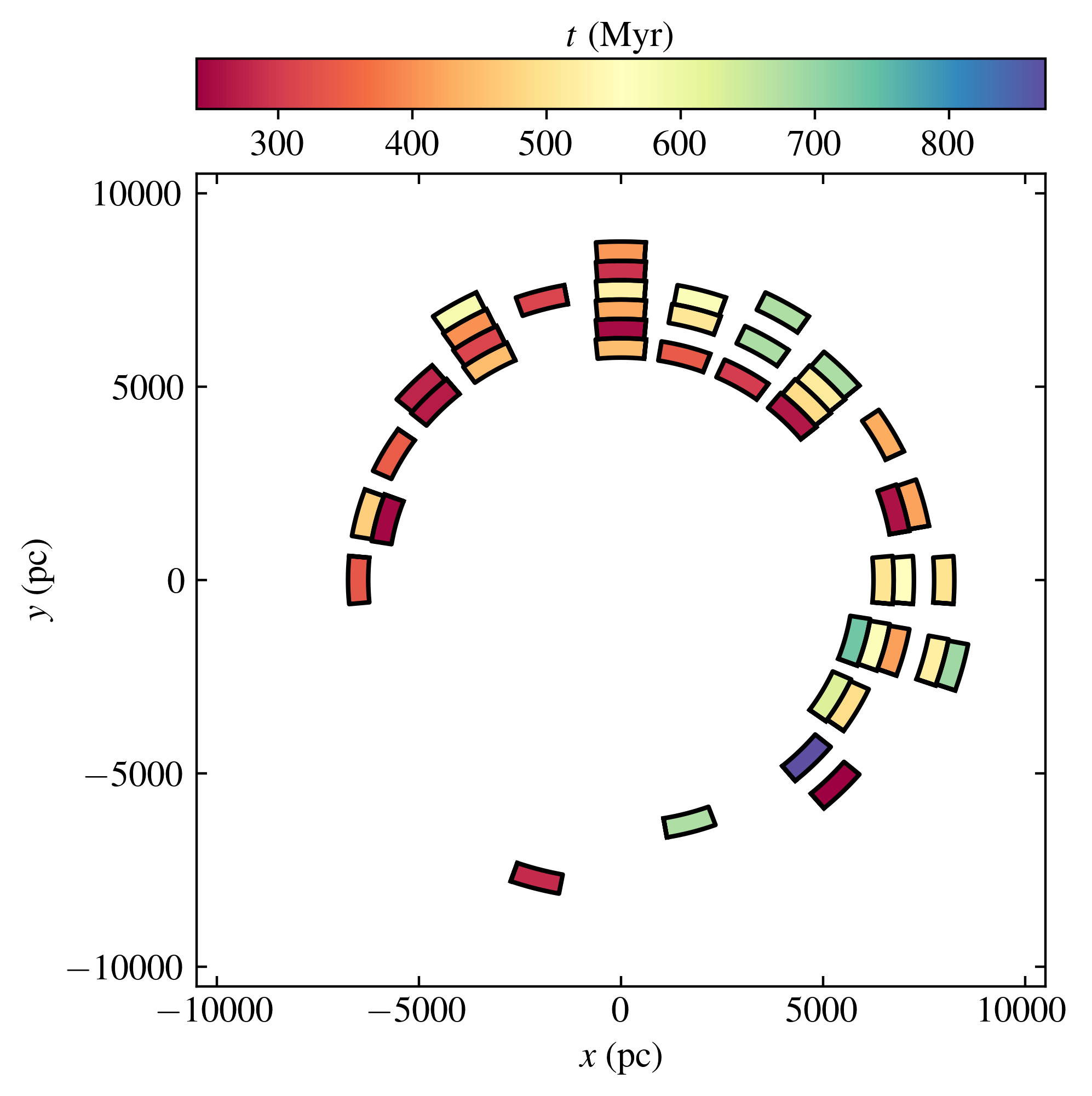}
    \caption{Spatial distribution of the inferred parameter $t$, for the case when the gravitational potential is free to vary in shape.}
    \label{fig:volumes_with_t_pert}
\end{figure}

\begin{figure}
	\includegraphics[width=1.\columnwidth]{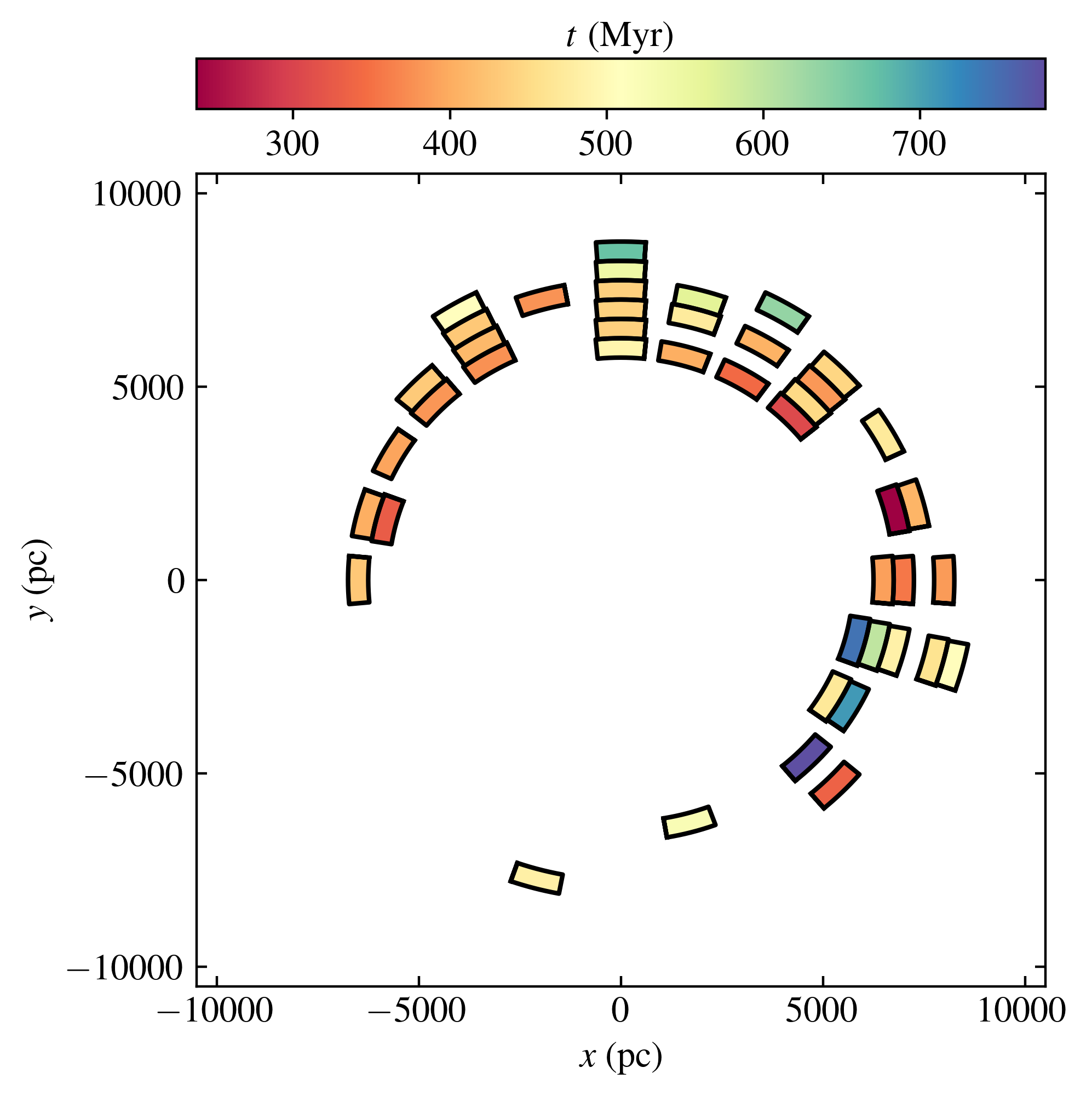}
    \caption{Spatial distribution of the inferred parameter $t$, for the case when the gravitational potential is fixed to its true shape and free to vary only in terms of its normalisation.}
    \label{fig:volumes_with_t_pert_fixed_shape}
\end{figure}

In this appendix we include a few supplementary plots. In Figs.~\ref{fig:hist_rel_err_800pc} and \ref{fig:hist_rel_err_800pc}, we show the relative errors for the inferred quantity $\Phi(800~\pc)$, as a histogram and in terms of its distribution in the disk plane. In Figs.~\ref{fig:spiral_7500_30}--\ref{fig:spiral_8500_90}, we show the gravitational potential, data histogram, data residual with respect to the fitted bulk, and the fitted spiral for six different data samples. The first is included as an example of a highly asymmetric gravitational potential, as a result of its proximity to the perturbing satellite. Finally, in Figs.~\ref{fig:volumes_with_t_pert} and \ref{fig:volumes_with_t_pert_fixed_shape}, we show how the inferred time $t$ is distributed in the disk plane, for the case when the shape of the vertical gravitational potential is either free to vary or fixed to its true shape in the simulation snapshot.

\begin{figure}
	\includegraphics[width=1.\columnwidth]{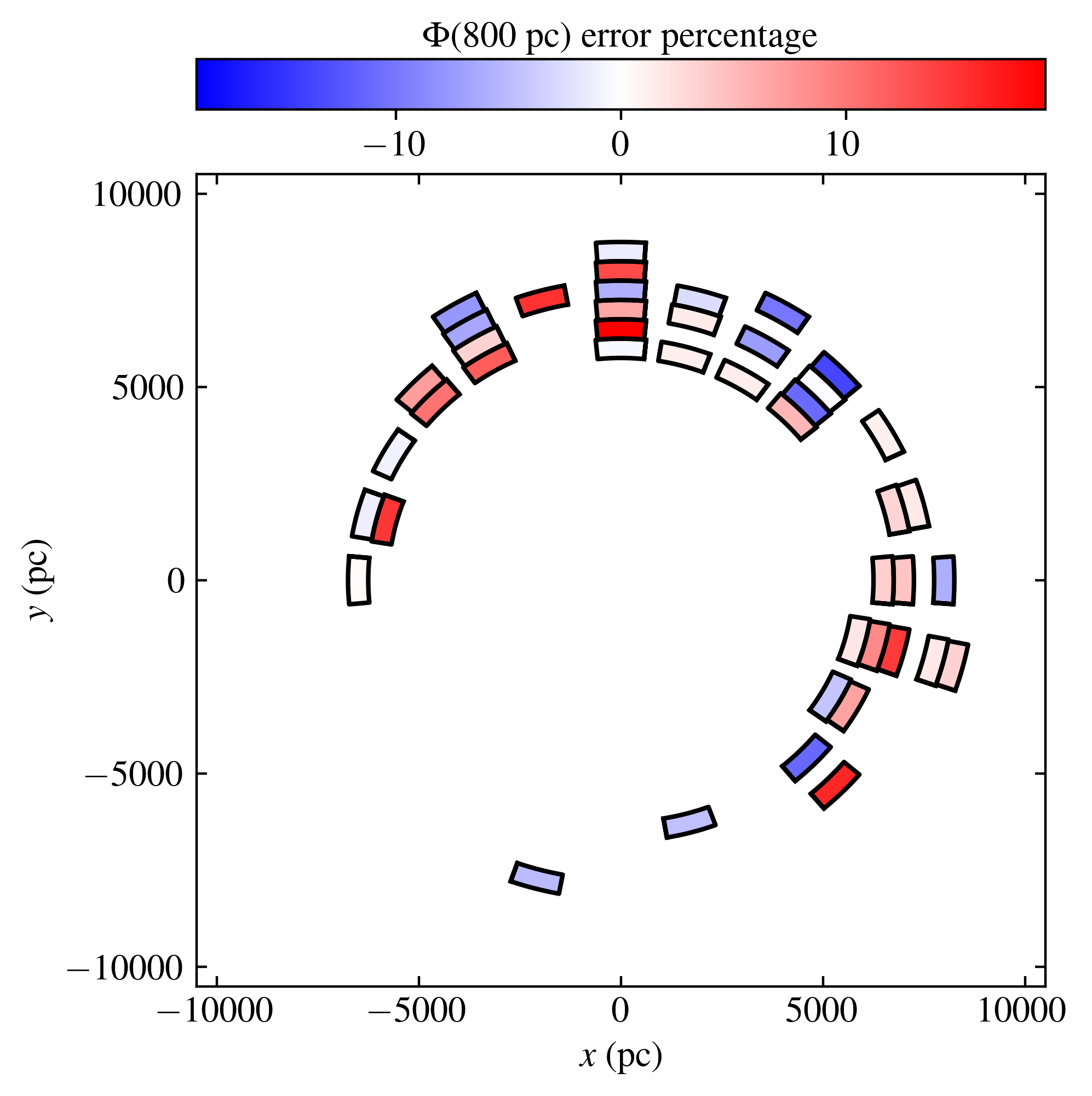}
    \caption{Relative errors of our spiral method for the quantity $\Phi(800~\pc)$.}
    \label{fig:volumes_rel_err_800pc}
\end{figure}

\end{appendix}

\end{document}